\documentclass[12pt,preprint,iop,apj]{emulateapj}  

\usepackage{graphicx}
\usepackage{txfonts}
\newcommand{\va}{c_{\mathrm{A}}}
\newcommand{\vai}{c_{\mathrm{Ai}}}

\newcommand{\csn}{c_{\rm sn}}
\newcommand{\csi}{c_{\rm sie}}

\newcommand{\cs}{c_{\rm s}}

\newcommand{\cti}{c_{\rm Tie}}

\newcommand{\rhoi}{\rho_{\rm i}}

\begin{document}

	\title{Kelvin-Helmholtz instability in partially ionized compressible plasmas}

	\shorttitle{KHI IN PARTIALLY IONIZED PLASMAS}

\shortauthors{R. Soler et al.}

\author{R. Soler$^{1}$, A. J. D\'iaz$^{2,3}$, J. L. Ballester$^4$, \& M. Goossens$^1$}
   \affil{$^1$Centre for Plasma Astrophysics, Department of Mathematics, Katholieke Universiteit Leuven,
              Celestijnenlaan 200B, 3001 Leuven, Belgium}
              \email{roberto.soler@wis.kuleuven.be}  

\affil{$^2$ Instituto de Astrof\'isica de Canarias, E-38200 La Laguna, Tenerife, Spain}

\affil{$^3$ Departamento de Astrof\'isica, Universidad de La Laguna, E-38206 La Laguna, Tenerife, Spain}

 \affil{$^4$Solar Physics Group, Departament de F\'isica, Universitat de les Illes Balears,
              E-07122, Palma de Mallorca, Spain}

\begin{abstract}
The Kelvin-Helmholtz Instability (KHI) has been observed in the solar atmosphere. Ion-neutral collisions may play a relevant role for the growth rate and evolution of the KHI in solar partially ionized plasmas as in, e.g.,  solar prominences. Here, we investigate the linear phase of the KHI at an interface between two partially ionized magnetized plasmas in the presence of a shear flow.  The effects of ion-neutral collisions and compressibility are included in the analysis. We obtain the dispersion relation of the linear modes and perform parametric studies of the unstable solutions. We find that in the incompressible case the KHI is present for any velocity shear regardless the value of the collision frequency. In the compressible case, the domain of instability depends strongly on the plasma parameters, specially the collision frequency and the density contrast. For high collision frequencies and low density contrasts the KHI is present for super-Alfv\'enic velocity shear only. For high density contrasts the threshold velocity shear can be reduced to sub-Alfv\'enic values. For the particular case of  turbulent plumes in prominences, we conclude that sub-Alfv\'enic flow velocities can trigger the KHI thanks to the ion-neutral coupling.

\end{abstract}


%________________________________________________________________

\section{Introduction}

The Kelvin-Helmholtz Instability (KHI) is a well-known magnetohydrodynamic instability that arises at the interface between two fluids in relative motion \citep[see the classical textbooks by][]{chandra,drazin}. The KHI is believed to operate on many astrophysical plasmas as, e.g., the magnetopause \citep[e.g.][]{hasegawa}, planetary magnetospheres \citep[e.g.,][]{nagano,miura84}, Earth's aurora  \citep[e.g.,][]{farrugia}, cometary tails \citep[e.g.,][]{comets}, protoplanetary disks \citep[e.g.,][]{gomez}, jets and outflows \citep[e.g.,][]{keppens99,keppens06}, among many other situations.

The KHI is also at work in the solar atmosphere. Recent observations by \citet{foullon} and  \citet{ofman} confirmed the presence of the KHI in the solar corona. Although the KHI has been observed  for the first time in the corona recently, there  is a large number of theoretical papers on the KHI in solar plasmas in the literature \citep[e.g.,][]{rae,andries1,andries2,holzwarth,terradas,solerKHI,temuryKHI,diaz}. More relevant for the present investigation are the observations by \citet{berger08,berger10,berger11} and \citet{ryutova} of turbulent flows and instabilities in solar prominences, which have been interpreted in terms of the Rayleigh-Taylor instability and the KHI. The peculiar properties of prominences, i.e., relatively cool and dense plasma condensations supported in the corona by the magnetic field, make them a very interesting subject for the investigation of the KHI. In particular, the fact that the prominence plasma is only partially ionized can have important effects on the behavior and evolution of the KHI.

It is known that  a velocity shear at the interface between two incompressible plasmas is always unstable in the absence of a magnetic field. The presence of a magnetic field component along the flow direction suppresses the KHI for sub-Alfv\'enic velocity shear in fully ionized plasmas \citep[see, e.g., ][]{chandra}. However, the situation is more complicated if the plasma is partially ionized as happens, e.g., in prominences. In a partially ionized plasma composed of ions and neutrals, the two species behave very differently when a magnetic field is applied. Neutrals are insensitive to the magnetic field so that their natural tendency is to be unstable for any velocity shear, while ions feel the stabilizing presence of the magnetic field. The coupling between neutrals and ions through collisions is therefore crucial for the growth rate of the KHI and its evolution \citep[see a discussion on the relevance of ion-neutral interaction for plasma instabilities in][]{lehnert}.

The KHI in partially ionized incompressible plasmas has been studied under different contexts \citep[e.g.,][]{chhajlani, birk, watson, shadmehri1,shadmehri2, kunz,jones}. The paper that we take as a reference for the present investigation is that by \citet{watson}. These authors studied the KHI due to shear flow at the interface between two incompressible and partially ionized plasmas. They applied their results to  weakly ionized outflows from massive stars interacting with the ambient interstellar material. They concluded that ion-neutral collisions cannot suppress the instability of neutrals for sub-Alfv\'enic velocity shear. Thus, a partially ionized incompressible plasma is unstable for any velocity shear.

\citet{watson} did not take the effect of compressibility into account. In fully ionized plasmas,  compressibility can have either a stabilizing or destabilizing role depending on the density contrast and the orientation of the flow with respect to the magnetic field direction. For a flow parallel to the magnetic field, compressibility has a stabilizing effect for small density contrast and large velocity shear, and a destabilizing effect for large density contrast and small velocity shear \citep[see, e.g.,][]{fejer,sen,gerwin,miura82,rae}. In partially ionized plasmas, the role of compressibility for the KHI was studied by \citet{prialnik} and \citet{comets} in the context of comet ionopauses. These authors took the flow of neutrals across the interface and did not study the case in which the flow of neutrals is along the magnetic field direction. To our knowledge, the joint effect of compressibility and ion-neutral collisions on the KHI has not been studied when the flow of both ions and neutrals is along the magnetic field.

In this paper we investigate the linear phase of the KHI at a sharp interface between two partially ionized plasmas in the presence of a shear flow. The flow direction of both ions and neutrals is along the magnetic field. Our study focuses on the effects of compressibility and ion-neutral collisions. In a particular application, we consider typical values of prominence plasma parameters. 

This paper is organized as follows. Section~\ref{sec:basic} contains the description of the equilibrium configuration and the basic equations. In Section~\ref{sec:collisionless} we study the KHI at an interface between two plasmas in the collisionless case, i.e., ion-neutral collisions are neglected. Later, we investigate in Section~\ref{sec:res:multi} the effect of ion-neutral collisions on the growth rates. Section~\ref{sec:appli} contains an application to  solar prominence plasmas. Finally, our conclusions are given in Section~\ref{sec:con}.

\section{Equilibrium and Basic Equations}
\label{sec:basic}

\subsection{Equilibrium configuration}

Our equilibrium is composed of  two partially ionized  and magnetized homogeneous plasmas separated by a sharp interface. We consider a hydrogen plasma composed of ions (protons), electrons, and neutral atoms. We use Cartesian coordinates. The $x$-direction is normal to the interface, so that the plane $x=0$ coincides with the interface. We denote by the indices 1 and 2 the regions corresponding to $x < 0$ and $x > 0$, respectively.  In the two regions,  the equilibrium magnetic field is straight and along the $z$-direction, namely ${\bf B} = B \hat{e}_z$, with $B$ constant. We denote by $B_1$ and $B_2$ the magnetic field strength in the two regions. The equilibrium plasma parameters are homogeneous and constant in both regions and the condition of pressure balance is satisfied at the interface. 

In the equilibrium the plasma species flow along the magnetic field with a constant flow velocity. We consider that all the species in the plasma flow with the same velocity. The equilibrium flow is ${\bf V}_1 = V_1 \hat{e}_z$ and ${\bf V}_2 = V_2 \hat{e}_z$. We denote by $V_1$ and $V_2$ the  flow velocities in regions 1 and 2, respectively. The flow velocities can be different on both sides of the interface, i.e., $V_1 \ne V_2$. Hence, there is a velocity shear that can trigger the KHI.

\subsection{Governing equations}

Plasma dynamics are governed by the momentum equation, the energy equation, and the mass conservation equation of ions, electrons, and neutrals. These equations contain terms accounting for collisions between the  species. In the two-fluid treatment of partially ionized plasmas \citep[see, e.g.,][]{brag,temury}, ions and electrons are considered together as an ion-electron fluid, while neutrals form another fluid that interacts with the ion-electron fluid by means of collisions. The generalized Ohm's Law along with the induction equation governing the evolution of the magnetic field are obtained by neglecting the electron inertia. In the remaining of this paper the indices i, e, and n refer to ions, electrons, and neutrals, respectively.

 In this investigation we restrict ourselves to the study of linear perturbations superimposed on the equilibrium state. Hence, the governing equations are linearized. The set of coupled differential equations governing linear perturbations from the equilibrium state are \citep[see, e.g.,][]{temury}
\begin{eqnarray}
\rho_{\rm i} \left( \frac{\partial {\bf v}_{\rm i}}{\partial t} + {\bf V} \cdot \nabla {\bf v}_{\rm i} \right) &=& - \nabla p_{\rm ie} + \frac{1}{\mu} \left( \nabla \times {\bf b} \right) \times {\bf B} \nonumber \\ &-& \alpha_{\rm in} \left(  {\bf v}_{\rm i} - {\bf v}_{\rm n}\right), \label{eq:momlinion} \\
\rho_{\rm n} \left( \frac{\partial {\bf v}_{\rm n}}{\partial t} + {\bf V} \cdot \nabla {\bf v}_{\rm n} \right) &=& - \nabla p_{\rm n} - \alpha_{\rm in} \left(  {\bf v}_{\rm n} - {\bf v}_{\rm i}\right),  \label{eq:momlinneu} \\
\frac{\partial {\bf b}}{\partial t} - \nabla \times \left( {\bf V} \times {\bf b} \right) &=& \nabla \times \left( {\bf v}_{\rm i} \times {\bf B} \right),  \label{eq:inductionlin} \\
\frac{\partial p_{\rm ie}}{\partial t} + {\bf V} \cdot \nabla p_{\rm ie} &=& - \gamma P_{\rm ie} \nabla \cdot  {\bf v}_{\rm i},  \label{eq:presslinion} \\
\frac{\partial p_{\rm n}}{\partial t} + {\bf V} \cdot \nabla p_{\rm n} &=& - \gamma P_{\rm n} \nabla \cdot  {\bf v}_{\rm n},  \label{eq:presslin} 
\end{eqnarray}
where $\bf V$ is the equilibrium flow velocity, ${\bf v}_{\rm i} = (v_{{\rm i} x},v_{{\rm i} y},v_{{\rm i} z})$ and ${\bf v}_{\rm n} = (v_{{\rm n} x},v_{{\rm n} y},v_{{\rm n} z})$ are the components of the velocity perturbation of ions and neutrals, respectively, $p_{\rm ie}$ and $p_{\rm n}$ are the pressure perturbations of the ion-electron and neutral fluids, respectively, ${\bf b} = (b_x, b_y, b_z)$ are the components of  the magnetic field perturbation,  $\rho_{\rm i}$ and $\rho_{\rm n}$ are the equilibrium densities of ions and neutrals, respectively, $P_{\rm ie}$ and $P_{\rm n}$ are the equilibrium gas pressure of the ion-electron and neutral fluids, respectively,  $\alpha_{\rm in}$ is the ion-neutral friction coefficient, $\mu$ is the magnetic permittivity, and $\gamma$ is the adiabatic index.

Equations~(\ref{eq:momlinion}) and (\ref{eq:momlinneu}) are the linearized momentum equations of the ion-electron fluid and neutrals, respectively, Equation~(\ref{eq:inductionlin}) is the linearized induction equation, and Equations~(\ref{eq:presslinion}) and (\ref{eq:presslin}) are the linearized energy equations of the ion-electron fluid and neutrals, respectively. In this equations we have performed some simplifications. Collisions of electrons with ions and neutrals are neglected because of the small momentum of electrons. Adiabatic perturbations are assumed and nonadiabatic mechanisms are omitted in Equations~(\ref{eq:presslinion}) and (\ref{eq:presslin}). In addition, magnetic diffusion terms are neglected in Equation~(\ref{eq:inductionlin}). These simplifications enable us to study the KHI analytically. The effect of the omitted mechanisms may be investigated in future works by means of numerical methods.

We express the ion-neutral friction coefficient, $\alpha_{\rm in}$,  as
\begin{equation}
 \alpha_{\rm in} = \rho_{\rm i} \rho_{\rm n} \gamma_{\rm in}, \label{eq:friction}
\end{equation}
where $\gamma_{\rm in}$ is the ion-neutral collision rate coefficient per unit mass. The friction coefficient vanishes in both the fully ionized and fully neutral cases. Instead of using $\gamma_{\rm in}$, in the remaining of this paper we use the collision frequency, which has a more practical physical meaning. Thus we define the ion-neutral, $\nu_{\rm in}$, and neutral-ion, $\nu_{\rm ni}$, collision frequencies as
\begin{equation}
 \nu_{\rm in} = \rho_{\rm i} \gamma_{\rm in}, \qquad \nu_{\rm ni} = \rho_{\rm n} \gamma_{\rm in}.
\end{equation}
Since $\nu_{\rm in}$ and $\nu_{\rm ni}$ are related by $\rho_{\rm n} \nu_{\rm in} = \rho_{\rm i}  \nu_{\rm ni}$, we use $\nu_{\rm in}$ in all the following expressions for simplicity \citep[see][for a discussion on the meaning of the different collision frequencies]{temuryhelium}. For our following analysis it is useful to define the total density as $\rho = \rho_{\rm i} + \rho_{\rm n}$, where we neglect the contribution of electrons. We define the neutral, $\zeta_{\rm n}$, and ion, $\zeta_{\rm i}$, fractions as $\zeta_{\rm n} = \rho_{\rm n} / \rho$ and $\zeta_{\rm i} = \rho_{\rm i} / \rho$, with $\zeta_{\rm n}  + \zeta_{\rm i} =1$.  The parameter $\zeta_{\rm n}$ is used here to indicate the plasma ionization degree. This parameter ranges from $\zeta_{\rm n} = 0$ for a fully ionized plasma to $\zeta_{\rm n} = 1$ for a neutral gas.

\subsection{Normal mode analysis}

We adopt an approach based on normal modes. We put the temporal dependence of the perturbations proportional to $\exp \left( - i \omega t \right)$, with $\omega$ the frequency. As the equilibrium is homogeneous in the $y$- and $z$-directions, we can write all perturbations proportional to $\exp \left( i k_y y + i k_z z \right)$, with $k_y$ and $k_z$ the wavenumbers in $y$- and $z$-directions, respectively. Due to the effect of the flow, the frequency is shifted so that $\Omega = \omega - k_z V$ is the Doppler-shifted frequency. The Lagrangian displacements of ions, ${\bf \xi}_{{\rm i}}$, and neutrals, ${\bf \xi}_{{\rm n}}$, are \citep[see, e.g.,][]{goossens92}
\begin{equation}
 {\bf \xi}_{{\rm i}} = \frac{i}{\Omega} {\bf v}_{{\rm i}}, \qquad {\bf \xi}_{{\rm n}} = \frac{i}{\Omega} {\bf v}_{{\rm n}}.
\end{equation}
The condition of continuity of ${\bf \xi}_{{\rm i}}$ and ${\bf \xi}_{{\rm n}}$ at $x=0$ imposes that the values of $k_y$ and $k_z$ are the same on both sides of the interface.

Since we consider compressible perturbations, we use  as our variables $\Delta_{\rm i} = \nabla \cdot {\bf \xi}_{{\rm i}}$ and $\Delta_{\rm n} = \nabla \cdot {\bf \xi}_{{\rm n}}$, i.e.,  the divergence of the ion and neutral displacements, respectively. Then, we can combine Equations~(\ref{eq:momlinion})-(\ref{eq:presslin}) and arrive at the two following coupled equations,
\begin{eqnarray}
\frac{\partial^2 \Delta_{\rm i}}{\partial x^2} - k_{\perp \rm ie}^2 \Delta_{\rm i} &=& i \mathcal{V}^2 \frac{ \rho_{\rm i} \nu_{\rm in}}{\gamma P_{\rm ie} + B^2/ \mu} \left( \frac{\partial^2 \Delta_{\rm n}}{\partial x^2}  -k_{\rm c}^2  \Delta_{\rm n}\right), \label{eq:basic1} \\
\frac{\partial^2 \Delta_{\rm n}}{\partial x^2} - k_{\perp \rm n}^2 \Delta_{\rm n} &=& i \Omega \frac{\rho_{\rm n} \nu_{\rm in}}{\gamma P_{\rm n}} \Delta_{\rm i}, \label{eq:basic2}
\end{eqnarray}
with   $\chi =  \rho_{\rm n} /\rhoi$ and $k_{\perp \rm i}^2$, $k_{\perp \rm n}^2$, $k^2_{\rm c}$, and $\mathcal{V}^2$  given by
\begin{eqnarray}
k_{\perp \rm ie}^2 &=& k_y^2 + k_z^2 - \frac{\Omega^2}{\vai^2 + \csi^2} \frac{ \Omega \left( \Omega + i \chi \nu_{\rm in} \right) }{ \Omega^2 - k_z^2 \tilde{c}_{\rm Tie}^2 },  \label{eq:kperpionsfull} \\
k_{\perp \rm n}^2 &=& k_y^2 + k_z^2 - \frac{ \Omega \left( \Omega + i  \nu_{\rm in} \right)}{\csn^2}, \label{eq:kperpneutralsfull} \\
k_{\rm c}^2 &=& k_y^2 + k_z^2 -  \frac{\Omega^2}{k_z^2 \vai^2 \csn^2} \Omega \left( \Omega + i (1+\chi) \nu_{\rm in} \right) , \\
\mathcal{V}^2 &=& \chi \frac{  k^2_z \csn^2}{\Omega^2  - k_z^2 \tilde{c}_{\rm Tie}^2 } \frac{\tilde{c}_{\rm Ai}^2}{ \Omega + i \nu_{\rm in}}, \\
\tilde{c}_{\rm Ai}^2 &=& \vai^2 \frac{ \Omega + i \nu_{\rm in}}{ \Omega + i (1+\chi) \nu_{\rm in} }, \quad \tilde{c}_{\rm Tie}^2 = \cti^2 \frac{ \Omega + i \nu_{\rm in}}{ \Omega + i (1+\chi) \nu_{\rm in} }, 
\end{eqnarray}
where $\vai^2 = B^2/\mu \rho_{\rm i}$ is the square of the Alfv\'en velocity of ions, $\csi^2 = \gamma P_{\rm ie}/\rho_{\rm i}$ and $\csn^2 = \gamma P_{\rm n}/\rho_{\rm n}$ are the square of the sound velocities of the ion-electron and neutral fluids, respectively, and $\cti^2 = \vai^2 \csi^2 / (\vai^2 + \csi^2)$ is the square of the cusp (or tube) velocity of the ion-electron fluid. The coupled Equations~(\ref{eq:basic1}) and (\ref{eq:basic2}) are the basic equations of the present investigation. In these equations, subscripts 1 or 2 must be used as appropriate when the equations are applied in the two regions of the equilibrium.

In the following sections we express the results in dimensionless units. We express velocities in units of $c_{\rm min} = \min \left( {\csn}_1 , {\csn}_2 \right)$ and distances in units of the typical length-scale in the $z$-direction, $L_z$. In particular applications  $L_z$ may be related to the wavelength of the plasma perturbations so that $L_z$ may be estimated from observations. Alternatively, if observations of wavelengths are not available, $L_z$ may be defined using, e.g., the typical scale of the variation of the plasma parameters along the $z$-direction. Thus, the dimensionless KHI growth rate, $\gamma_{\rm KHI}$, is computed as
\begin{equation}
 \gamma_{\rm KHI} =  \frac{ \omega_{\rm I} L_z}{c_{\rm min}},
\end{equation}
where $\omega_{\rm I}$ is the imaginary part of the frequency of the unstable solution. The dimensionless velocity shear at the interface, $\Delta v$, is 
\begin{equation}
 \Delta v = \frac{ \left| V_1 - V_2  \right| }{c_{\rm min}}.
\end{equation}
Unless otherwise stated, we use this nondimensionalization in the remaining of this paper.

\section{Collisionless Plasma}

\label{sec:collisionless}

First, we investigate the KHI when collisions between ions and neutrals are neglected.  This preliminary study will enable us to understand better the results when the coupling between ions and neutrals through collisions is included. By setting $\nu_{\rm in}=0$, Equations~(\ref{eq:basic1}) and (\ref{eq:basic2}) become,
\begin{eqnarray}
\frac{\partial^2 \Delta_{{\rm i}}}{\partial x^2} - k_{\perp \rm ie}^2 \Delta_{{\rm i}} &=& 0,  \label{eq:basic1d} \\
\frac{\partial^2 \Delta_{{\rm n}}}{\partial x^2} - k_{\perp \rm n}^2 \Delta_{{\rm n}} &=& 0. \label{eq:basic2d},
\end{eqnarray}
with $k_{\perp \rm ie}^2$ and $k_{\perp \rm n}^2$ now given by
\begin{eqnarray}
k_{\perp \rm ie}^2 &=& k_y^2 + \frac{\left( k_z^2 \csi^2 - \Omega^2 \right)\left( k_z^2 \vai^2 - \Omega^2 \right)}{\left( \csi^2 + \vai^2 \right) \left( k_z^2 \cti^2 - \Omega^2 \right)}, \label{eq:kisimple} \\
k_{\perp \rm n}^2 &=& k_y^2 +\frac{k_z^2 \csn^2 - \Omega^2}{\csn^2}. \label{eq:knsimple}
\end{eqnarray}
In the absence of ion-neutral collisions Equation~(\ref{eq:basic1d}) governs the ion-electron fluid and Equation~(\ref{eq:basic2d}) governs neutrals. We study these two equations separately.

\subsection{Instability of neutrals}

The solutions to Equation~(\ref{eq:basic2d}) for perturbations vanishing at $x \to \pm \infty$ are
\begin{equation}
\Delta_{{\rm n}} = \left\{ 
\begin{array}{lll}
A_{\rm n1} \exp( k_{\rm \perp n1} x) & \textrm{if} & x < 0,\\
A_{\rm n2} \exp( -k_{\rm \perp n2} x) & \textrm{if} & x > 0,
\end{array}
 \right.
\end{equation} 
where $A_{\rm n1}$ and $A_{\rm n2}$ are constants. To obtain the dispersion relation we impose the continuity of  the normal Lagrangian displacement and the neutral pressure perturbation at $x=0$. The dispersion relation is
\begin{equation}
\rho_{\rm n1}  \Omega_1^2 + \rho_{\rm n2}  \Omega_2^2 \mathcal{F}_{\rm n} = 0, \label{eq:disperneu}
\end{equation}
with $\mathcal{F}_{\rm n} = k_{\rm \perp n1} / k_{\rm \perp n2}$ the compressibility factor of neutrals.

Let us consider first the incompressible limit, i.e., ${\csn}_{1} \to \infty$ and ${\csn}_{2} \to \infty$ so that $\mathcal{F}_{\rm n} \to 1$. The analytical solution of Equation~(\ref{eq:disperneu}) in the incompressible case is
\begin{equation}
\omega = \frac{\rho_{\rm n1} V_1 + \rho_{\rm n2} V_2}{\rho_{\rm n1} + \rho_{\rm n2}} k_z \pm i k_z \left| V_1 - V_2 \right| \frac{\sqrt{\rho_{\rm n1}  \rho_{\rm n2}}}{\rho_{\rm n1} + \rho_{\rm n2}}. \label{eq:solneu}
\end{equation}
Note that Equation~(\ref{eq:solneu}) is independent of $k_y$. The first term on the right-hand side of Equation~(\ref{eq:solneu}) is the real part of $\omega$, which corresponds to the Doppler shift due to the flow. The second term is the imaginary part of $\omega$ and informs us about the stability of the solution. The solution with the $+$ sign is a growing, unstable mode, while the solution with the $-$ sign is a damped disturbance. The unstable solution is present for any value of the velocity shear. This is the classical incompressible hydrodynamic KHI \citep[see, e.g.][]{chandra,drazin}. 

\begin{figure}[!t]
\centering
 \includegraphics[width=0.75\columnwidth]{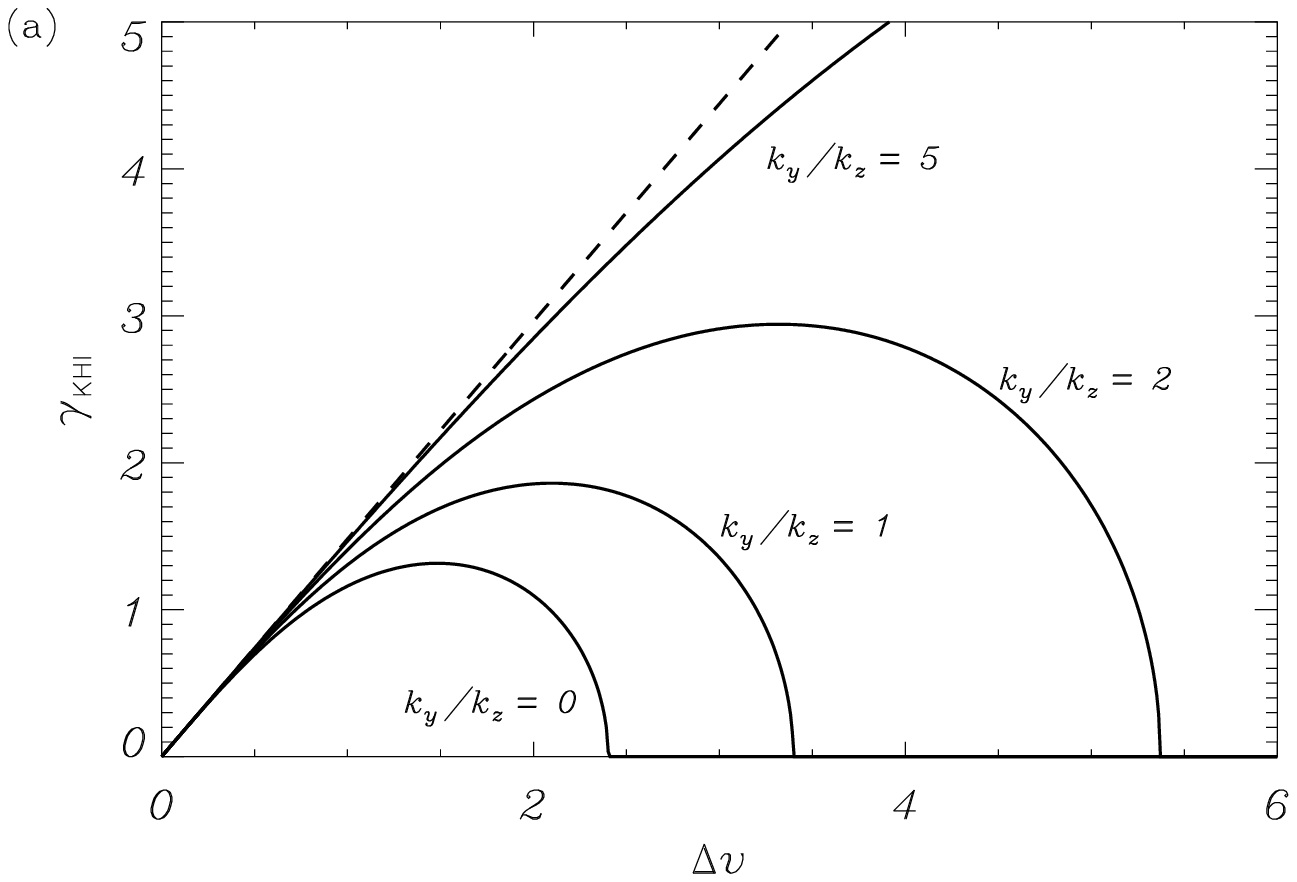}
\includegraphics[width=0.75\columnwidth]{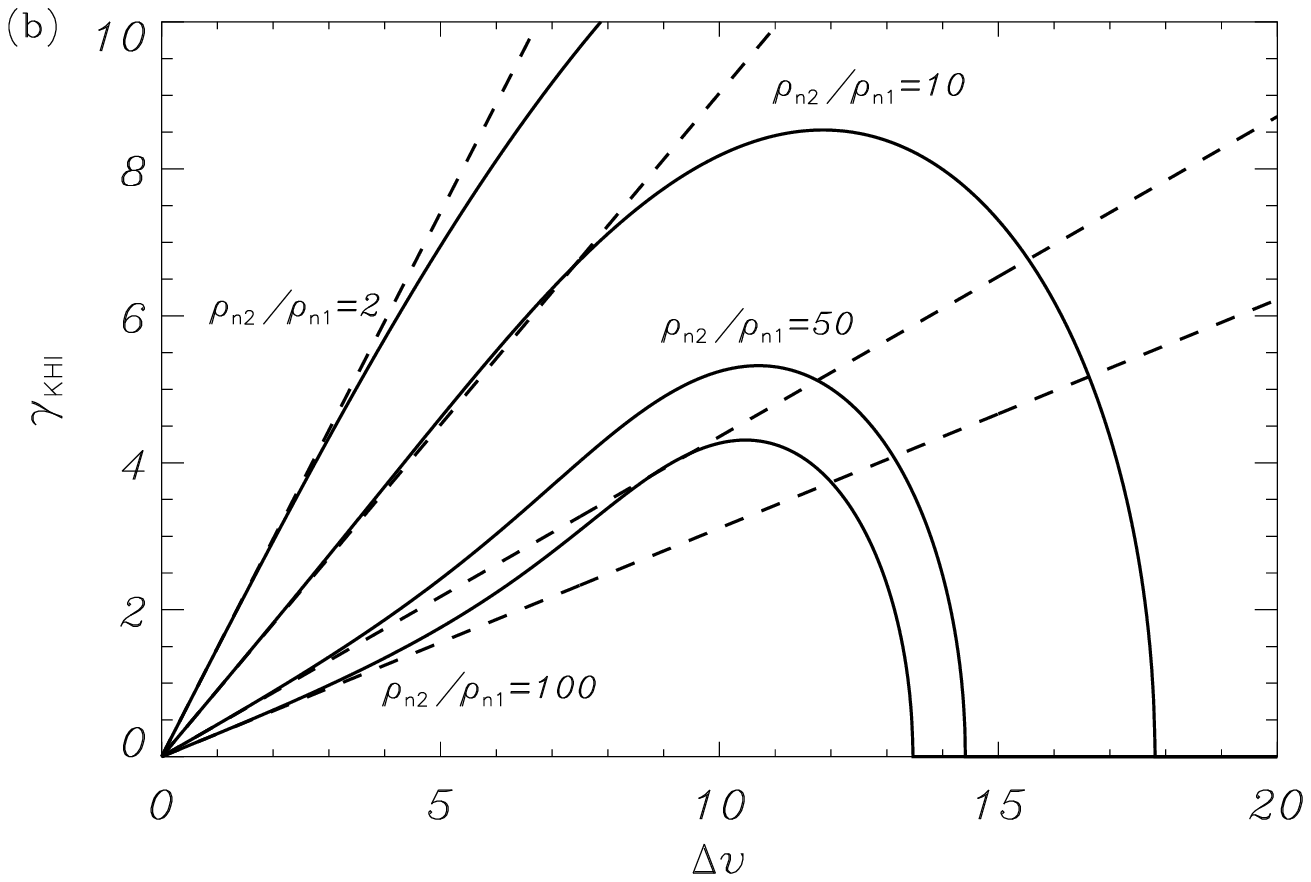}
\caption{Results for collisionless neutrals. Dimensionless growth rate, $\gamma_{\rm KHI}$, as a function of the dimensionless velocity shear at the interface, $\Delta v$. (a) Results for $\rho_{\rm n2}/\rho_{\rm n1}=2$ and different values of $k_y / k_z$. (b) Results for $k_y / k_z = 10$ and different values of $\rho_{\rm n2}/\rho_{\rm n1}$. The  dashed lines in both panels correspond to the incompressible case. In all cases $k_z L_z = \pi$. \label{fig:neutrals}}
\end{figure}

Now we study the compressible case. We solve Equation~(\ref{eq:disperneu}) numerically using dimensionless units. First we consider the case of low density contrast. We fix $\rho_{\rm n2}/\rho_{\rm n1}=2$ and compute the growth rate as a function of $\Delta v$ for different values of $k_y / k_z$ (see Figure~\ref{fig:neutrals}(a)). We compare these results with that in the incompressible limit given by Equation~(\ref{eq:solneu}) (dashed line in Figure~\ref{fig:neutrals}(a)). The ratio $k_y / k_z$ controls the effect of compressibility. For $k_y / k_z \to \infty$ both compressible and incompressible results agree. This means that compressibility plays no role when the wave vector in perpendicular to the flow direction. For finite $k_y / k_z$ compressibility is able to suppress the KHI for velocity shear larger than a critical value.  Compressibility cannot completely suppress the KHI for small velocity shear. In the particular case $k_y = 0$ the critical velocity shear remains supersonic. 

Next we set $k_y / k_z = 10$, so that compressibility has a small effect. In Figure~\ref{fig:neutrals}(b) we display $\gamma_{\rm KHI}$ versus $\Delta v$ for different values of the density contrast, $\rho_{\rm n2}/\rho_{\rm n1}$. The critical velocity shear for stabilization gets reduced as the density contrast increases. We find that for small shear and large density contrast the growth rate in the compressible case is larger than the incompressible value. Hence, when the effect of compressibility is small and the density contrast is large, compressibility destabilizes for small velocity shear and stabilizes for large velocity shear. These results agree with early investigations on the compressible hydrodynamic KHI \citep[see, e.g.,][]{fejer,sen,gerwin,miura82}.  

\subsection{Instability of the ion-electron fluid}

\begin{figure}[!t]
\centering
  \includegraphics[width=0.75\columnwidth]{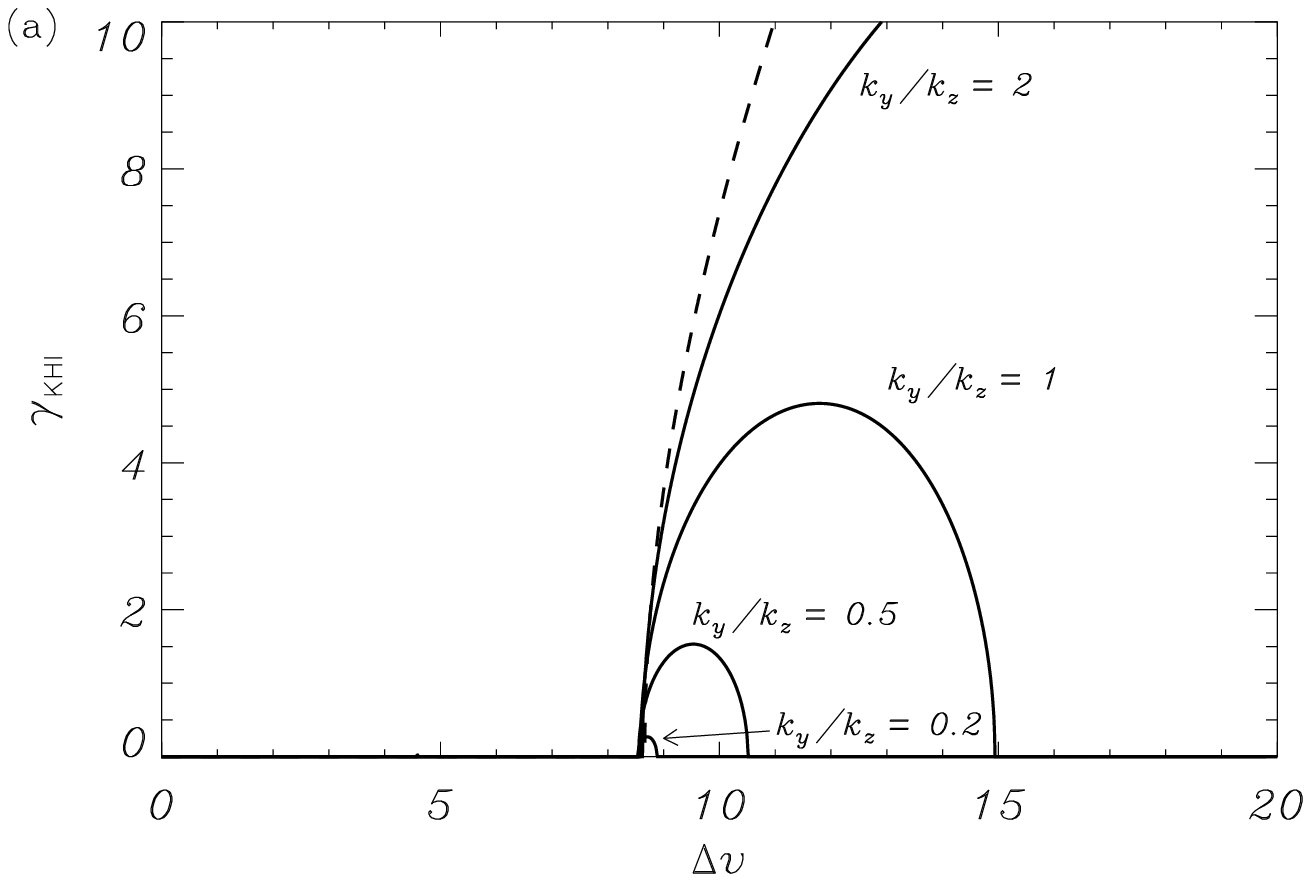}
\includegraphics[width=0.75\columnwidth]{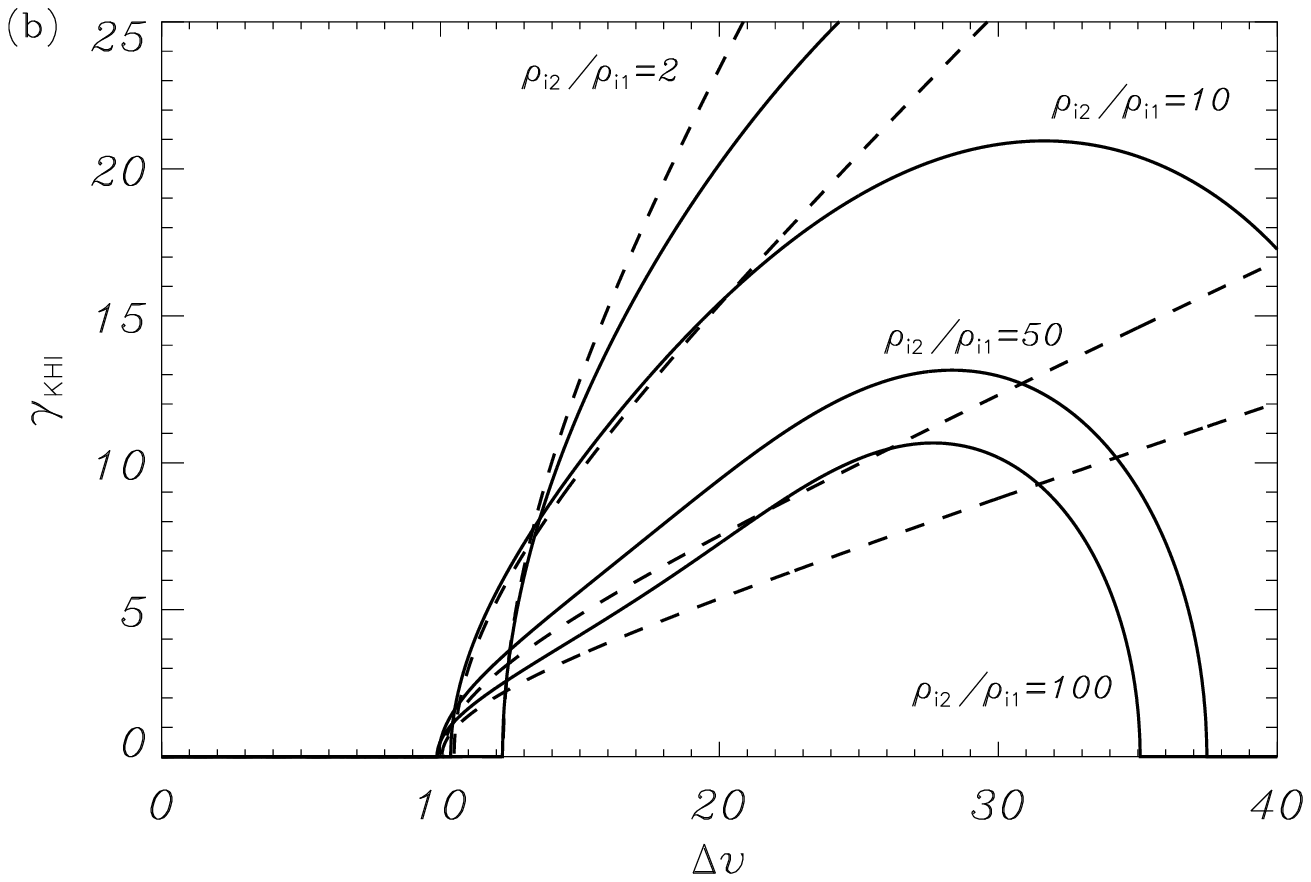}
\caption{Results for the collisionless ion-electron fluid. Dimensionless growth rate, $\gamma_{\rm KHI}$, as a function of the dimensionless velocity shear at the interface, $\Delta v$. (a) Results for $\rho_{\rm i2}/\rho_{\rm i1}=2$ and different values of $k_y / k_z$. (b) Results for $k_y / k_z = 5$ and different values of $\rho_{\rm i2}/\rho_{\rm i1}$. The  dashed lines in both panels correspond to the incompressible case.  An all computations, $\vai / \csi = 5$ and $k_z L = \pi$. \label{fig:ions}}
\end{figure}

Now we study the instability of the ion-electron fluid. The solutions to Equation~(\ref{eq:basic1d})  for perturbations vanishing at $x \to \pm \infty$ are
\begin{equation}
\Delta_{{\rm i}} = \left\{ 
\begin{array}{lll}
A_{\rm i1} \exp( k_{\rm \perp i1} x) & \textrm{if} & x < 0,\\
A_{\rm i2} \exp( -k_{\rm \perp i2} x) & \textrm{if} & x > 0,
\end{array}
 \right.
\end{equation} 
where $A_{\rm i1}$ and $A_{\rm i2}$ are constants. As before, we impose the continuity of  the normal Lagrangian displacement and the total (gas plus magnetic) pressure perturbation at $x=0$. The dispersion relation is
\begin{equation}
\rho_{\rm i1} \left( \Omega_1^2 - k_z^2 {\vai^2}_1 \right) + \rho_{\rm i2} \left( \Omega_2^2 - k_z^2 {\vai^2}_2 \right) \mathcal{F}_{\rm i} = 0, \label{eq:disperion}
\end{equation}
with $\mathcal{F}_{\rm i} = k_{\rm \perp i1} / k_{\rm \perp i2}$ the compressibility factor of the ion-electron fluid.

We repeat the same process as for neutrals and study the incompressible case first, i.e., $\mathcal{F}_{\rm i} \to 1$. The solution to Equation~(\ref{eq:disperion}) in the incompressible case is
\begin{eqnarray}
\omega &=& \frac{\rho_{\rm i1} V_1 + \rho_{\rm i2} V_2}{\rho_{\rm i1} + \rho_{\rm i2}} k_z \nonumber \\ &\pm & k_z \left[ \frac{B_1^2 + B_2^2}{\mu \left( \rho_{\rm i1} + \rho_{\rm i2} \right)} - \left( V_1 - V_2 \right)^2 \frac{\rho_{\rm i1}  \rho_{\rm i2}}{\left( \rho_{\rm i1} + \rho_{\rm i2} \right)^2} \right]^{1/2}. \label{eq:solion}
\end{eqnarray}
Note that Equation~(\ref{eq:solion}) is independent of $k_y$. In the absence of flows, Equation~(\ref{eq:solion}) reduces to the frequency of the incompressible surface mode driven by magnetic tension, namely
\begin{equation}
\omega = \pm k_z \left( \frac{B_1^2 + B_2^2}{\mu \left( \rho_{\rm i1} + \rho_{\rm i2} \right)} \right)^{1/2} \equiv \omega_{\rm k}. \label{eq:omegak}
\end{equation}
In Equation~(\ref{eq:solion}) the second term on the right-hand side informs us about the stability of the solution. If the terms within the brackets become negative, the solution with the $+$ sign is a unstable mode, while the solution with the $-$ sign is a damped mode. This is the classical incompressible magnetohydrodynamic KHI \citep[see, e.g.][]{chandra,drazin}. The threshold velocity shear that triggers the KHI is
\begin{equation}
\left| V_1 - V_2 \right| > \left( \frac{\rho_{\rm i1} +\rho_{\rm i2} }{\rho_{\rm i1}\rho_{\rm i2} } \frac{B_1^2 + B_2^2}{\mu} \right)^{1/2}. \label{eq:criticalv}
\end{equation}
In the absence of magnetic field, $B_1 = B_2 = 0$ and the solution is unstable for any velocity shear. Thus, we recover the well-known result that the longitudinal component of the magnetic field has a stabilizing effect on the KHI \citep[see, e.g.,][]{chandra}.

Now we investigate the role of compressibility. For simplicity we have restricted ourselves to the case $\va > \cs$. First we fix the density contrast to $\rho_{\rm i2}/\rho_{\rm i1}=2$ and explore the solutions of Equation~(\ref{eq:disperion}) for different values of the ratio $k_y / k_z$ (see Figure~\ref{fig:ions}(a)). Incompressible and compressible results agree when the ratio $k_y / k_z$ is large. As the ratio $k_y / k_z$ decreases, perturbations are more affected by the stabilizing effect of compressibility. Compressibility can suppress the KHI for velocity shear larger than a critical value. Thus, the solution is unstable in a range of velocity shears in between the threshold value due to the magnetic field (Equation~(\ref{eq:criticalv})) and the critical value due to compressibility. For $k_y = 0$ the combined effect of compressibility and the magnetic field completely suppresses the KHI.

Again, we study the case in which compressibility has a small effect and the density contrast is large. To do so, we set $k_y / k_z = 5$ and compute the growth rate versus $\Delta v$ for different values of $\rho_{\rm i2}/\rho_{\rm i1}$ (see Figure~\ref{fig:ions}(b)). When the density contrast is large compressibility destabilizes for small velocity shear and stabilizes for large velocity shear in comparison with the incompressible limit. The behavior of the ion-electron fluid is therefore the same as that of neutrals regarding compressibility.

\section{Collisional plasma}

\label{sec:res:multi}

Here we consider the coupling between ions and neutrals through collisions. Our purpose is to assess the effect of ion-neutral collisions on the results of the collisionless case (Section~\ref{sec:collisionless}).   We express the collision frequency,  $\nu_{\rm in}$, in units of the surface wave frequency, $\omega_{\rm k}$, defined in Equation~(\ref{eq:omegak}). Thus, the ratio $\nu_{\rm in} / \omega_{\rm k}$ controls the relevance of ion-neutral collisions. The larger $\nu_{\rm in} / \omega_{\rm k}$, the more important the effect of collisions.

To find the solutions of the coupled Equations~(\ref{eq:basic1}) and (\ref{eq:basic2}), we use Equation~(\ref{eq:basic2}) to express  $\Delta_{\rm i}$ in terms of $\Delta_{{\rm n}}$ and its derivatives as
\begin{equation}
 \Delta_{\rm i} = - i \frac{1}{\Omega} \frac{\gamma P_{\rm n}}{\rho_{\rm n} \nu_{\rm in}} \left( \frac{\partial^2 \Delta_{\rm n}}{\partial x^2} - k_{\perp \rm n}^2 \Delta_{\rm n} \right).
\end{equation}
Next we use this last expression in Equation~(\ref{eq:basic1}) to obtain an equation for $\Delta_{{\rm n}}$ only, namely
\begin{equation}
 \frac{\partial^4 \Delta_{\rm n}}{\partial x^4} - \left( k_{\perp \rm n}^2 + k_{\perp \rm ie}^2 -k_\mathcal{V}^2 \right)  \frac{\partial^2 \Delta_{\rm n}}{\partial x^2} + \left( k_{\perp \rm n}^2 k_{\perp \rm ie}^2 -k_{\rm c}^2 k_\mathcal{V}^2 \right) \Delta_{\rm n} = 0, \label{eq:completelamn}
\end{equation}
with $k_\mathcal{V}^2$ defined as
\begin{equation}
 k_\mathcal{V}^2 = \Omega \mathcal{V}^2 \frac{\rho_{\rm i} \rho_{\rm n} \nu_{\rm in}^2}{\gamma P_{\rm n} \left( \gamma P_{\rm ie} +B^2/\mu \right)}.
\end{equation}
Equation~(\ref{eq:completelamn}) is a linear differential equation with constant coefficients. The solution to Equation~(\ref{eq:completelamn}) is of the form,
\begin{equation}
\Delta_{{\rm n}} = A \exp \left( k x \right), \label{eq:soltry}
\end{equation}
with $A$ an arbitrary constant and $k$ a wavenumber to be determined. By substituting this expression in Equation~(\ref{eq:completelamn}) we obtain an algebraic equation for $k^2$, whose solution is
\begin{eqnarray}
k^2_\pm &=& \frac{1}{2} \left( k_{\rm \perp ie}^2 + k_{\rm \perp n}^2 -k_\mathcal{V}^2  \right) \nonumber \\
 &\pm& \frac{1}{2} \left[ \left( k_{\rm \perp ie}^2 - k_{\rm \perp n}^2 \right)^2  + k_\mathcal{V}^2\left( k_\mathcal{V}^2 - 2 k_{\rm \perp ie}^2 - 2 k_{\rm \perp n}^2 + 4 k_{\rm c}^2  \right) \right]^{1/2}, \label{eq:kmasmenos}
\end{eqnarray}
Hence, two independent wavenumbers, namely $k_+^2$ and $k_-^2$, are possible. The general expression of $\Delta_{{\rm n}}$  must contain the contributions of both $k_+^2$ and $k_-^2$.

To understand why there are two different values of $k^2$ it is instructive assuming a weak coupling between the species, so that the quadratic terms in $\nu_{\rm in}$ can be neglected and $k_\mathcal{V}^2 \approx 0 $. Then the two independent wavenumbers in Equation~(\ref{eq:kmasmenos}) simplify to
\begin{equation}
k^2_+ \approx k_{\rm \perp ie}^2, \qquad k^2_- \approx k_{\rm \perp n}^2. 
\end{equation}
The two values of $k^2$ reduce to the wavenumbers of the ion-electron fluid and neutrals in the collisionless case, respectively (see Equations~(\ref{eq:kperpionsfull}) and (\ref{eq:kperpneutralsfull})). In the general case, $k_\mathcal{V}^2 \ne 0 $ and these two wavenumbers are coupled due to ion-neutral collisions. 

Taking into account the above expressions and the condition that perturbations must vanish at $x \to \pm \infty$, we write the general form for  $\Delta_{\rm n}$ on both sides of the interface as	
\begin{equation}
\Delta_{{\rm n}} = \left\{ 
\begin{array}{lll}
A_1^+ \exp( k_{\rm + 1} x) + A_1^- \exp( k_{\rm - 1} x) & \textrm{if} & x < 0,\\
A_2^+ \exp( -k_{\rm + 2} x) + A_2^- \exp(- k_{\rm - 2} x) & \textrm{if} & x > 0,
\end{array}
 \right. \label{eq:solgenxin}
\end{equation} 
where $A_1^+$, $A_1^-$, $A_2^+$, and $A_2^-$ are constants. We need to impose appropriate boundary conditions at $x=0$ to obtain the dispersion relation. The boundary conditions are the continuity of the normal Lagrangian displacement of ions and neutrals, the ion-electron total pressure perturbation, and the neutral total pressure perturbation. After expressing these quantities as functions of $\Delta_{{\rm n}}$ and its derivatives and applying the boundary conditions, we get an algebraic system of four equations for the constants $A_1^+$, $A_1^-$, $A_2^+$, and $A_2^-$. The nontrivial solution of the system provides us with the dispersion relation, i.e., the determinant formed by the coefficients of the equations set equal to zero. The expression of the dispersion relation is given in the Appendix~\ref{app}.

\subsection{Incompressible case}

Before embarking on the investigation of the unstable solutions in the compressible case, we first discuss the incompressible limit. This is the case studied by \citet{watson}. In the incompressible case, the dispersion relation is
\begin{eqnarray}
&& \left\{ \rho_{\rm i1} \left[ \Omega_1 \left( \Omega_1 + i \chi_1 \nu_{\rm in 1}  \right) - k_z^2 c_{\rm A i1}^2 \right] \right. \nonumber \\
 &+& \left.  \rho_{\rm i2} \left[ \Omega_2 \left( \Omega_2 + i \chi_2 \nu_{\rm in 2}  \right) - k_z^2 c_{\rm A i2}^2 \right] \right\}\nonumber \\
&\times & \left[ \rho_{\rm n1} \Omega_1 \left( \Omega_1 + i \nu_{\rm in 1}  \right) + \rho_{\rm n2} \Omega_2 \left( \Omega_2 + i \nu_{\rm in 2}  \right)  \right] \nonumber \\
&+& \left(  \rho_{\rm n1} \Omega_1 \nu_{\rm in 1} + \rho_{\rm n2} \Omega_2 \nu_{\rm in 2} \right)^2 = 0. \label{eq:dispmultiinc}
\end{eqnarray}
Using the same notation as \citet{watson}, our Equation~(\ref{eq:dispmultiinc}) reverts to their Equation~(26).

\begin{figure}[!t]
\centering
\includegraphics[width=0.75\columnwidth]{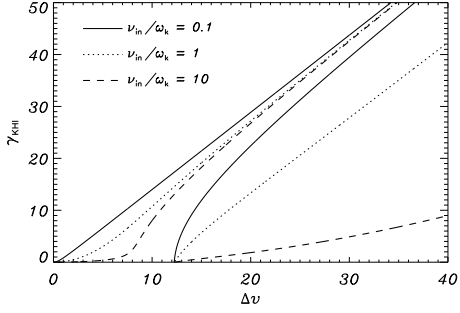}
\caption{Results in the incompressible case. Dimensionless growth rate, $\gamma_{\rm KHI}$, as a function of the dimensionless velocity shear at the interface, $\Delta v$, in a collisional plasma with $\nu_{\rm in1} / \omega_{\rm k} = \nu_{\rm in2} / \omega_{\rm k}=$ 0.1, 1, and 10. In all computations $\rho_{\rm 2}/\rho_{\rm 1}=2$, $\zeta_{\rm n1} = \zeta_{\rm n2} = 0.5$, and $k_z L_z = \pi$.  \label{fig:incomp}}
\end{figure}

 The solutions of Equation~(\ref{eq:dispmultiinc}) are plotted in Figure~\ref{fig:incomp}. We want to assess how the results in the collisionless case change when we increase the collision frequency. For this reason, we consider three different values of the ratio $\nu_{\rm in} / \omega_{\rm k}$. In the case $\nu_{\rm in} / \omega_{\rm k} \ll 1$ (see the solid lines corresponding to $\nu_{\rm in} / \omega_{\rm k} = 0.1$) we essentially recover the results of the collisionless case. There are two unstable solutions or branches, the first one related to neutrals and the second one related to the ion-electron fluid.  The growth rates of the two solutions are very similar to those indicated in Section~\ref{sec:collisionless} for the collisionless case. Ion-neutral collisions become important when  $\nu_{\rm in}$ and $\omega_{\rm k}$ are of the same order (see the dotted line corresponding to $\nu_{\rm in} / \omega_{\rm k} = 1$). Due to collisions, the growth rate of the two unstable solutions decrease. For $\nu_{\rm in} / \omega_{\rm k} \gg 1$ (see the dashed line corresponding to $\nu_{\rm in} / \omega_{\rm k} = 10$) there is a strong coupling between the two fluids, so that the two unstable branches are affected by the behavior of both fluids. The growth rate of the first unstable branch is much larger than that of the second branch when $\nu_{\rm in} / \omega_{\rm k} \gg 1$.

\begin{figure}[!t]
\centering
\includegraphics[width=0.75\columnwidth]{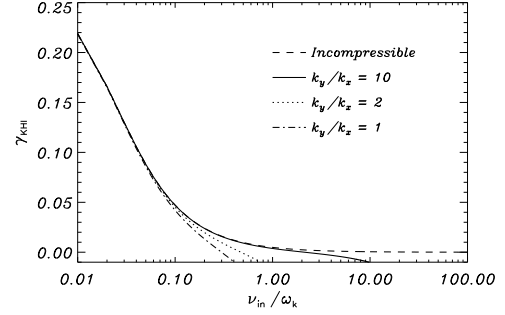}
\caption{$\gamma_{\rm KHI}$ as a function of  $\nu_{\rm in} / \omega_{\rm k}$, in a collisional plasma with $\Delta v = 0.2$  for $\nu_{\rm in1} = \nu_{\rm in2} = \nu_{\rm in}$, $\rho_{\rm 2}/\rho_{\rm 1}=2$, $\vai / \csi =5$, $\csi =\csn$, $\zeta_{\rm n1} = \zeta_{\rm n2} = 0.5$, and $k_z L = \pi$. The meaning of the different lines is indicated within the Figure. \label{fig:multinu}}
\end{figure}

Importantly, we find that in the incompressible case the first unstable branch is unstable for any velocity shear regardless the value of $\nu_{\rm in} / \omega_{\rm k}$. The growth rate for small velocity shear continuously decreases when $\nu_{\rm in} / \omega_{\rm k}$ increases, but stabilization is never achieved. Figure~\ref{fig:multinu} displays the growth rate of the first unstable branch as a function of $\nu_{\rm in} / \omega_{\rm k}$ for small velocity shear ($\Delta v  = 0.2$). In Figure~\ref{fig:multinu} the dashed line corresponds to the incompressible case. As for small shear the solution is governed by neutrals, this  means that ion-neutral collisions are not able to stabilize neutrals even in the limit $\nu_{\rm in} / \omega_{\rm k} \gg 1$.  Here our results are consistent with those of \citet{watson}, who found that the growth rate of the solution related to neutrals is approximately proportional to $\nu_{\rm in}^{-1}$ \citep[see details in the Appendix of][]{watson}.

\begin{figure}[!t]
\centering
\includegraphics[width=0.75\columnwidth]{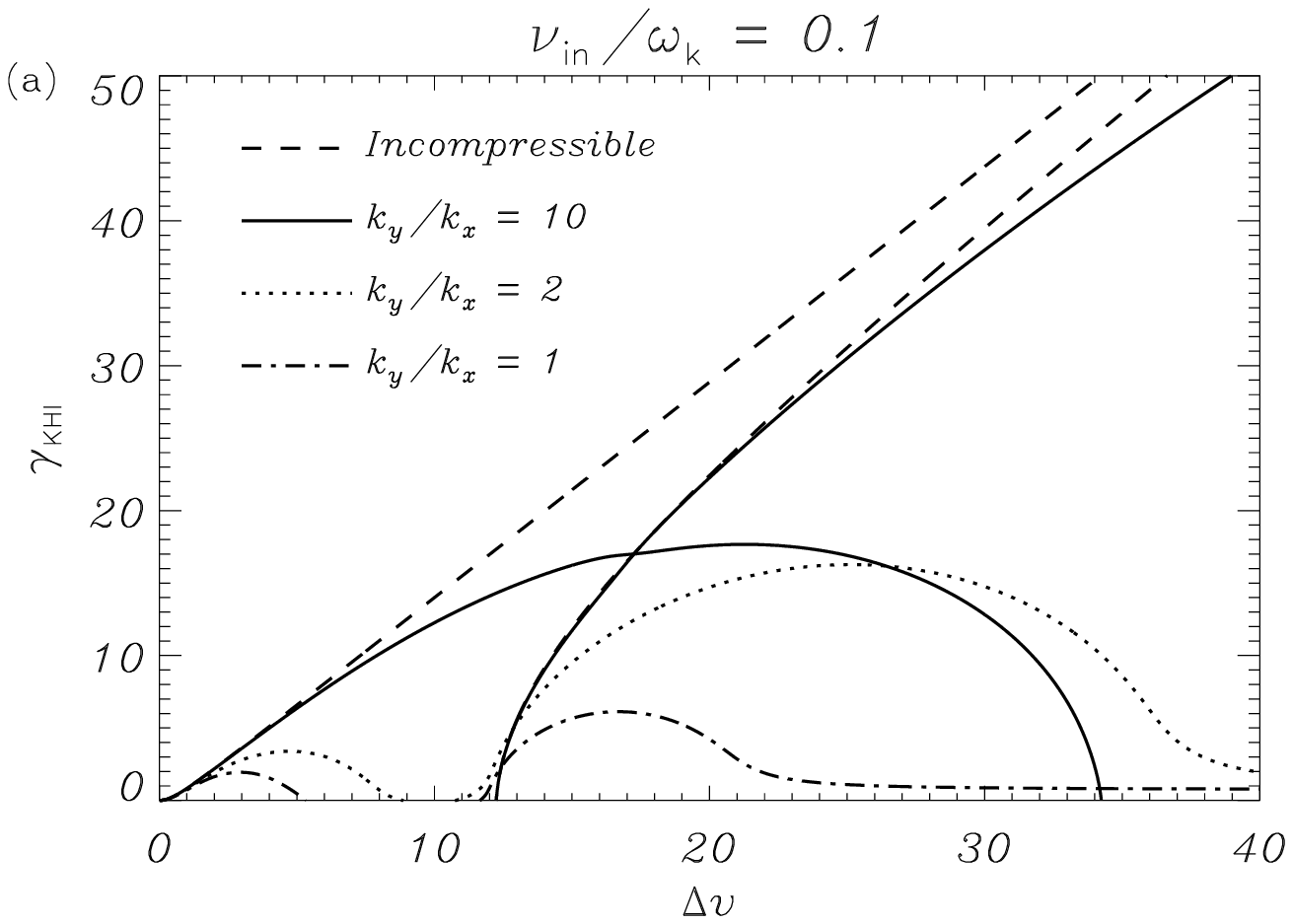}
\includegraphics[width=0.75\columnwidth]{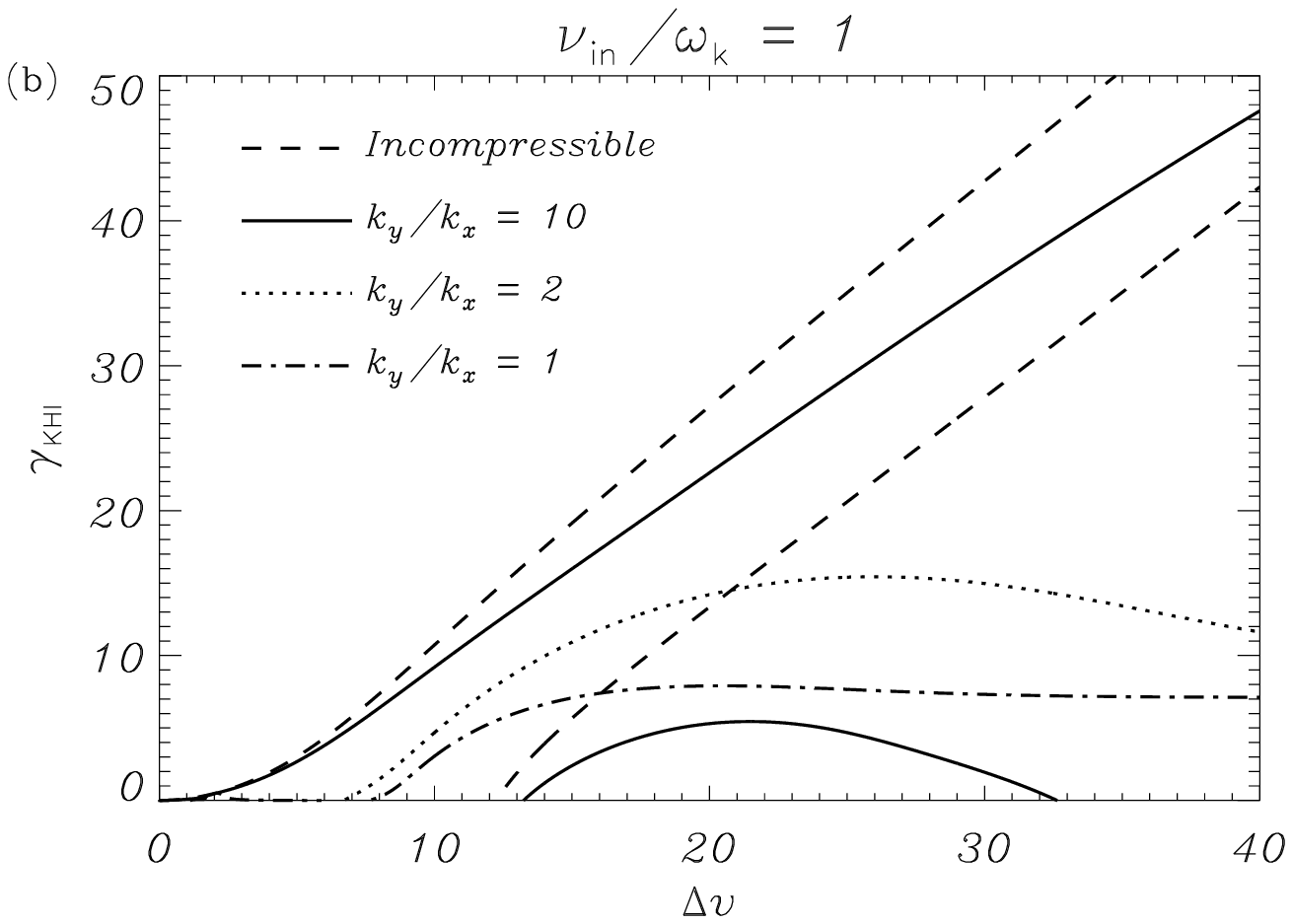}
\includegraphics[width=0.75\columnwidth]{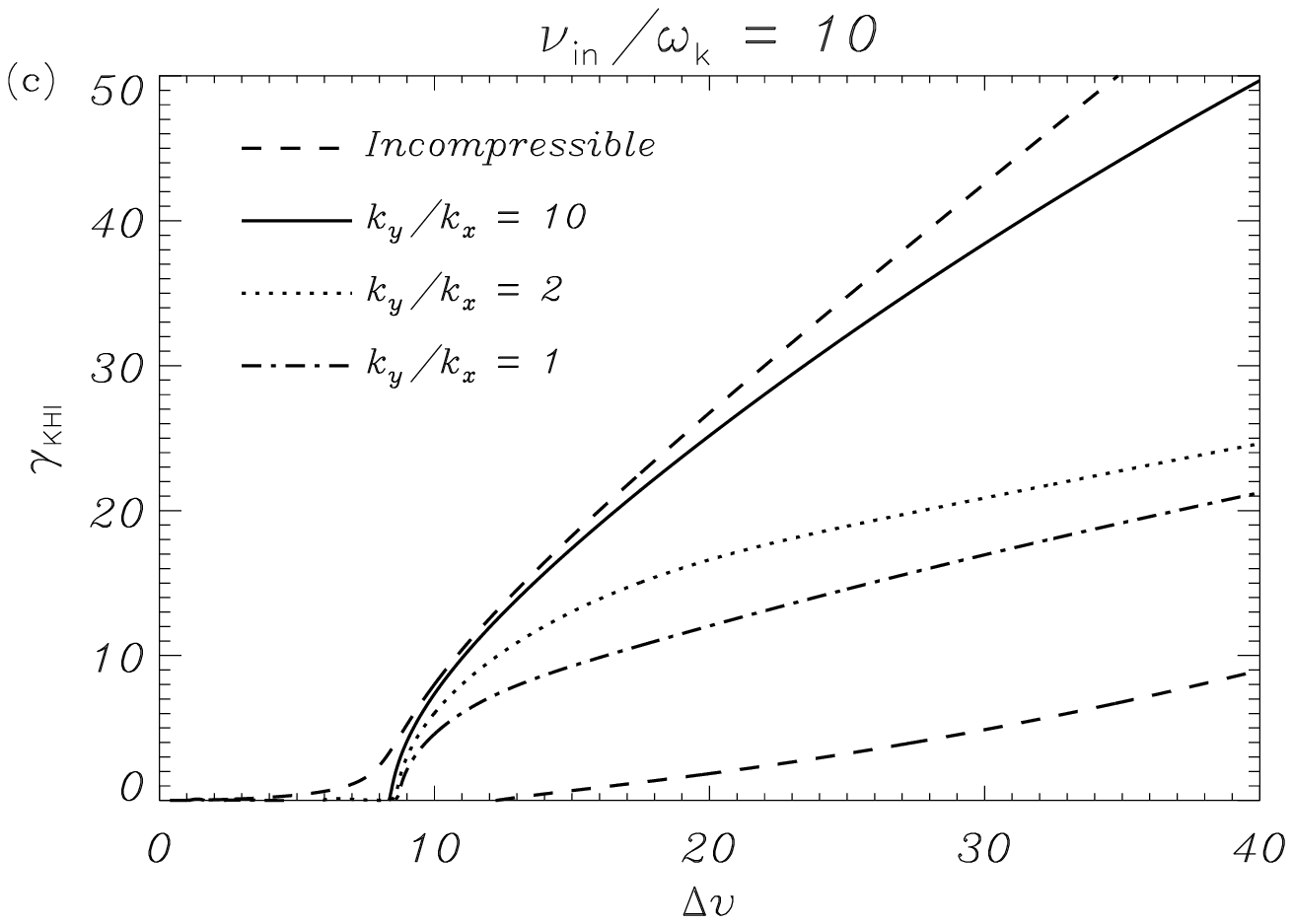}
\caption{Dimensionless growth rate, $\gamma_{\rm KHI}$, as a function of the dimensionless velocity shear at the interface, $\Delta v$, in a collisional plasma with (a) $\nu_{\rm in1} / \omega_{\rm k} = \nu_{\rm in2} / \omega_{\rm k}= 0.1$, (b) $\nu_{\rm in1} / \omega_{\rm k} = \nu_{\rm in2} / \omega_{\rm k}= 1$, and (c) $\nu_{\rm in1} / \omega_{\rm k} = \nu_{\rm in2} / \omega_{\rm k}= 10$. The meaning of the different lines is indicated within the panels. In all computations $\rho_{\rm 2}/\rho_{\rm 1}=2$, $\vai / \csi =5$, $\csi =\csn$, $\zeta_{\rm n1} = \zeta_{\rm n2} = 0.5$, and $k_z L_z = \pi$.  \label{fig:multi}}
\end{figure}

\subsection{Compressible case}
 
\label{sec:comp}

Here we discuss how the results in the incompressible limit are affected by compressibility. Compressibility was not considered by \citet{watson}. For simplicity, we restrict ourselves to the case $\va > \cs$. We compare the results when we increase  $\nu_{\rm in} / \omega_{\rm k}$. In the panels of Figure~\ref{fig:multi} we plot the results for three different values of $\nu_{\rm in} / \omega_{\rm k}$ with a density contrast of $\rho_2 / \rho_1 = 2$. Different line styles are used depending on the value of  $k_y / k_z$. For comparison purposes we have also plotted using dashed lines the equivalent growth rates of the incompressible case. We obtain that for $k_y / k_z \gg 1$ the growth rates of the incompressible limit are recovered. The results displayed in Figure~\ref{fig:multi} show a complicated evolution of the unstable modes when $\nu_{\rm in} / \omega_{\rm k}$ is increased. In the following paragraphs we summarize the main results contained in Figure~\ref{fig:multi}.

When $\nu_{\rm in} / \omega_{\rm k} \ll 1$ (see Figure~\ref{fig:multi}a for $\nu_{\rm in} / \omega_{\rm k}=0.1$) we obtain two unstable branches as in the incompressible case. Now the two branches are affected by compressibility. Compressibility causes the solutions to be closer to the stability threshold, i.e., their growth rates are smaller than in the incompressible case.  However, compressibility does not result in the complete stabilization of all solutions. In particular the solutions with large $k_y / k_z$ cannot be completely stabilized by compressibility. Only the solutions with small $k_y / k_z$ can be stabilized by compressibility for large enough $\Delta v$.

When $\nu_{\rm in}$ and $\omega_{\rm k}$ are of the same order (see Figure~\ref{fig:multi}b for $\nu_{\rm in} / \omega_{\rm k}=1$) two unstable branches are present when $k_y / k_z \gg 1$ only (see the solid lines in Figure~\ref{fig:multi}b corresponding to $k_y / k_z =10$). However, as $k_y / k_z$ decreases the second branch disappears and only the first branch remains. The behavior of the first branch is complex because it  has two regions of instability. The first one takes place for small velocity shear and is not visible at the scale of Figure~\ref{fig:multi}b because it has very small growth rates. The second region of instability is the solution visible in Figure~\ref{fig:multi}b and has much larger growth rates. It is only present for velocity shear larger than a threshold shear. The threshold shear becomes super-Alfv\'enic when $k_y / k_z$ decreases. This is an important difference compared to the incompressible case.

Finally, when $\nu_{\rm in} / \omega_{\rm k} \gg 1$ (see Figure~\ref{fig:multi}c for $\nu_{\rm in} / \omega_{\rm k}=10$) only one unstable branch is found. It has two regions of instability as before, but again the first one is not visible at the scale of the Figure~\ref{fig:multi}c. Eventually, the first unstable region disappears when $\nu_{\rm in} / \omega_{\rm k}$ is increased to a large enough value. The second region of instability remains regardless the value of  $\nu_{\rm in} / \omega_{\rm k}$  and has a threshold velocity shear. The threshold shear is super-Alfv\'enic and is approximately given by Equation~(\ref{eq:criticalv}) if the density of ions is replaced by the sum of the densities of ions and neutrals. Thus, in the limit $\nu_{\rm in} / \omega_{\rm k} \to \infty$ we obtain the behavior of a plasma where ions, electrons, and neutrals behave as a single-fluid. In this limit, the linear growth rate is the same as in the fully ionized case but replacing the density of ions by the sum of densities of ions and neutrals.

\begin{figure}[!t]
\centering
\includegraphics[width=0.75\columnwidth]{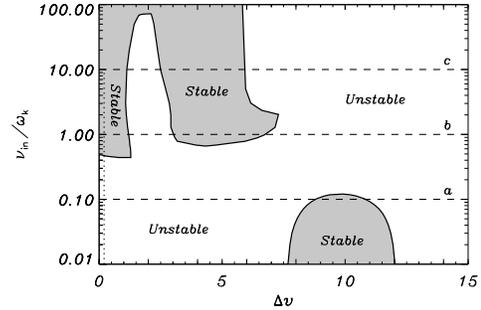}
\caption{Instability diagram $\nu_{\rm in} / \omega_{\rm k}$ vs. $\Delta v$ with $\rho_{\rm 2}/\rho_{\rm 1}=2$, $\vai / \csi =5$, $\csi =\csn$, $\zeta_{\rm n1} = \zeta_{\rm n2} = 0.5$, $k_y/k_z = 2$, and $k_z L = \pi$. The shaded areas denote regions of stability, while the white zone denotes instability. The three horizontal dashed lines correspond to the values of $\nu_{\rm in} / \omega_{\rm k}$ used in the three panels of Figure~\ref{fig:multi}. The vertical dotted line corresponds to the value of $\Delta v$ used in Figure~\ref{fig:multinu}. \label{fig:diagram}}
\end{figure}

The results discussed in the previous paragraphs are qualitatively displayed in Figure~\ref{fig:diagram}, which corresponds to the instability diagram $\nu_{\rm in} / \omega_{\rm k}$ vs. $\Delta v$ when the remaining parameters are kept constant. In the top left part of the Figure we can see a small region of instability between two stable zones. This is the region of instability with very small growth rates that is not visible in Figures~\ref{fig:multi}b,c. Eventually, for large enough $\nu_{\rm in} / \omega_{\rm k}$ this region of instability disappears and the two stable zones on both sides merge. For the range of parameters considered in Figure~\ref{fig:diagram} the two stable regions merge for $\nu_{\rm in} / \omega_{\rm k} \approx 70$. However we must remark that the critical value of $\nu_{\rm in} / \omega_{\rm k}$ depends strongly on the equilibrium parameters so that it may vary substantially when the parameters are changed. Hence we have to be cautious since it is not possible to give a general value for the ratio $\nu_{\rm in} / \omega_{\rm k}$ to achieve stabilzation in all cases.

In Figure~\ref{fig:diagram} we clearly see the transition between the collisionless case and the single-fluid case. Let us discuss this transition in physical terms. For $\nu_{\rm in} / \omega_{\rm k} \ll 1$ there is only one stable zone due to the combined effect of the magnetic field and compressibility. On the left-hand side of the stable zone the instability is governed by neutrals, while on the right-hand side the instability is mainly due to the ion-electron fluid. Thus, the behavior is the same as in the collisionless case. As $\nu_{\rm in} / \omega_{\rm k}$ increases the effect of collisions is, first, to destabilize the system so that the stable region is removed. Hence for a weak coupling between the species the stabilizing roles of compressibility and the magnetic field are overridden. This is mainly a consequence of the influence of neutrals, which are insentitive to the magnetic field, on ions. As  $\nu_{\rm in} / \omega_{\rm k}$ keeps increasing and the coupling between species gets stronger two additional stable regions appear at low velocity shear. This first stable region is due to the fact that ion-neutral collisions are able to stabilize neutrals for small velocity shear. The second stable zone appears due to the combined effect of collisions and the magnetic field on the ion-electron fluid. For $\nu_{\rm in} / \omega_{\rm k} \gg 1$ ion-neutral collisions can eventually suppress the unstable zone between the two stable regions so that single stable zone is present for low velocity shear as in the single-fluid limit.

Compressibility has a strong impact on the KHI in comparison to the incompressible case, specially when we depart from the limit $\nu_{\rm in} / \omega_{\rm k} \ll 1$. In general, the effect of compressibilty gets stronger as $k_y/k_z$ decreases. For large $k_y/k_z$ the effect of compressibility is weak and the growth rates tend to those in the incompressible case. Compressibility can suppress the KHI for small velocity shear so that a threshold velocity shear appears. We refer again to Figure~\ref{fig:multinu} where this result can be seen in more detail. Figure~\ref{fig:multinu} displays the growth rate of the first unstable branch as a function of $\nu_{\rm in} / \omega_{\rm k}$ for small velocity shear ($\Delta v  = 0.2$). The smaller $k_y / k_z$, and so the stronger the effect of compressibility, the smaller the value of $\nu_{\rm in} / \omega_{\rm k}$ needed for stabilization.

\begin{figure}[!t]
\centering
\includegraphics[width=0.75\columnwidth]{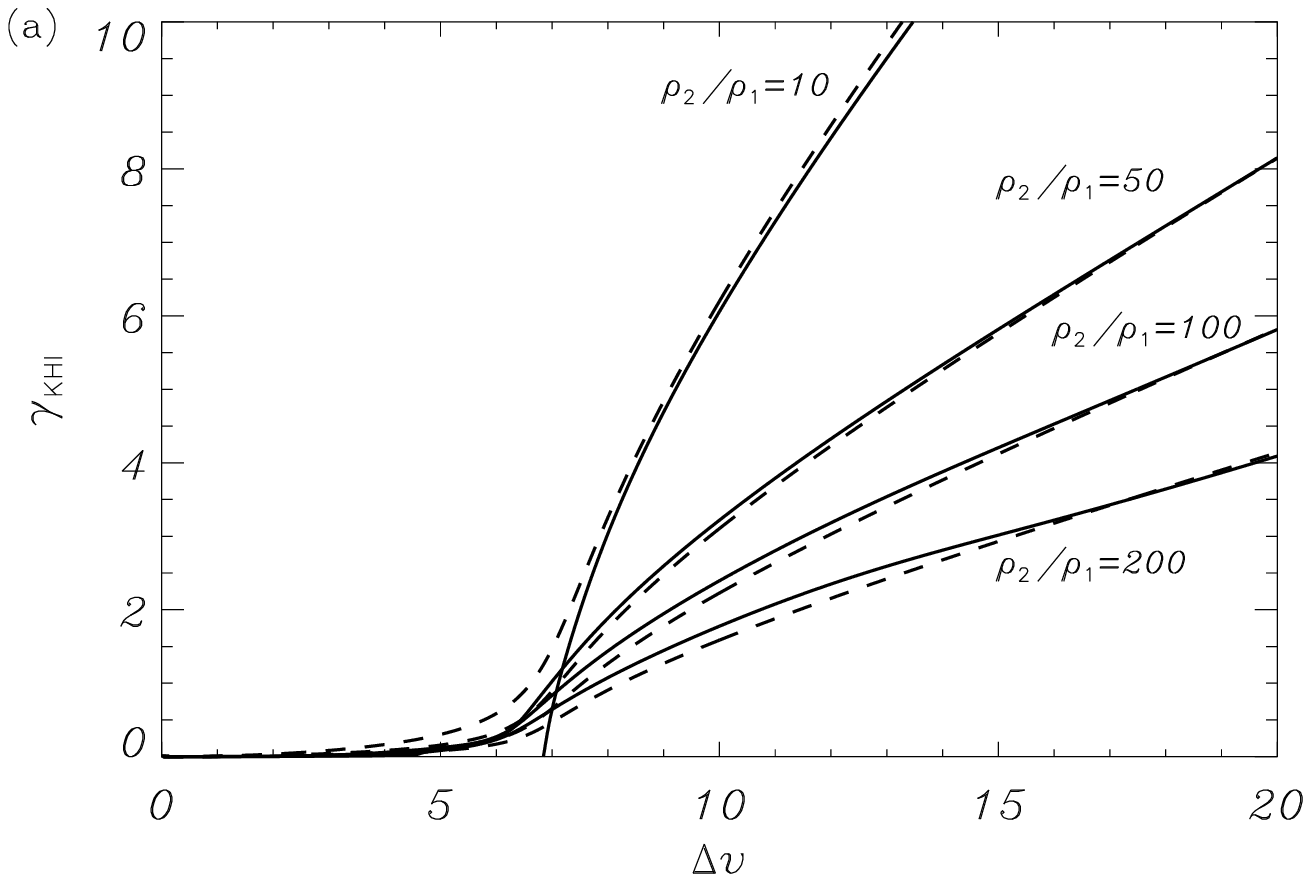}
\includegraphics[width=0.75\columnwidth]{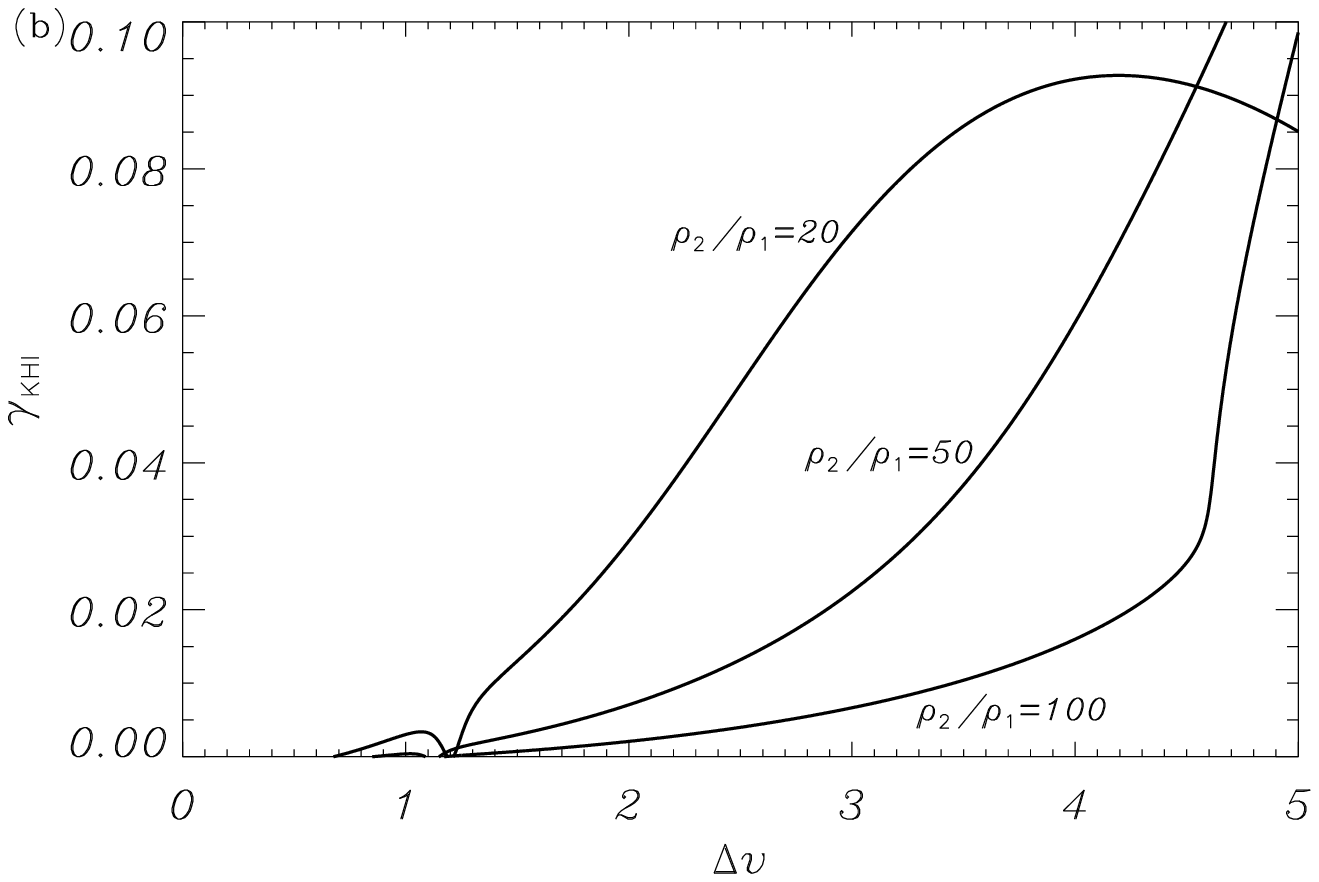}
\caption{(a) $\gamma_{\rm KHI}$, as a function of  $\Delta v$ with $\nu_{\rm in1} / \omega_{\rm k} = \nu_{\rm in2} /  \omega_{\rm k} = 10$ and different values of $\rho_{\rm 2}/\rho_{\rm 1}$. The remaining parameters are the same as in Figure~\ref{fig:multi}. (b) Same as panel (a) but for small $\Delta v$. The dashed lines in panel (a) correspond to the incompressible case. \label{fig:multicon}}
\end{figure}

Next we study the effect of increasing the density contrast, $\rho_2 / \rho_1$, when the effect of compressibility is small. We set $k_y / k_z = 10$ and $\nu_{\rm in} / \omega_{\rm k} = 10$ so that we focus on the case $\nu_{\rm in} / \omega_{\rm k} \gg 1$. Figure~\ref{fig:multicon}(a) displays $\gamma_{\rm KHI}$ versus  $\Delta v$ for different values of $\rho_2 / \rho_1$. At the scale of Figure~\ref{fig:multicon}(a), compressible and incompressible cases give very similar results. The differences are more important for small velocity shear. This can be seen in more detail in Figure~\ref{fig:multicon}(b), which focuses on small values of $\Delta v$. The results displayed in Figure~\ref{fig:multicon}(b) are not visible at the scale of Figure~\ref{fig:multicon}(a). For small velocity shear the growth rate has a complicated dependence on the density contrast. Importantly, we find that when the effect of compressibility is small and high density contrast the threshold shear can be reduced to sub-Alfv\'enic velocities.

 Although the case $\va > \cs$ is appropriate for solar atmospheric applications, the case $\cs > \va$ also deserves some attention. We have performed several additional computations in the case $\cs > \va$ and have assessed the effect of increasing the sound velocity. In summary, we find that when the sound velocity gets larger the results tend to those of the incompressible case. Thus, we need to take small values of the ratio $k_y / k_z$ for compressibility to have an impact on the growth rates. In the limit $\cs \to \infty$ the incompressible case is recovered and the growth rates are independent of $k_y / k_z$.

\section{Applications to solar prominences}
\label{sec:appli}

Here we apply our results to the case of flows in solar prominences. Prominences are cool and dense plasma condensations supported in the solar corona by magnetic forces. Due to the relatively low temperature of the prominence material ($\sim 10^4$~K), the prominence plasma is only partially ionized. Mass flows prominences have been frequently reported \citep[e.g.,][]{zirker98, wang1999, kucera2003, lin2003,okamoto,ahn2010}. Therefore, prominences are a good subject for an application of the theory developed in the present paper.

We perform two different applications. First, we consider the case of turbulent upflows in quiescent prominences observed by  \citet{berger08,berger10,berger11} and \citet{ryutova}. Later, we study flows along thin threads of prominence barbs. Here we use dimensional units. We take $L_y$ and $L_z$ as the typical length scales for the perturbations en the $y$ and $z$-directions, respectively.  We compute $k_y$ and $k_z$ as
\begin{equation}
 k_y = \frac{\pi}{L_y}, \qquad  k_z = \frac{\pi}{L_z}.
\end{equation}
The expression for the ion-neutral collision frequency, $\nu_{\rm in}$, in a hydrogen plasma is \citep[see, e.g.,][]{solerpartial}
\begin{equation}
 \nu_{\rm in} = \frac{\rhoi}{2 m_{\rm p}} \sqrt{\frac{16 k_{\rm B} T}{\pi m_{\rm p}}} \sigma_{\rm in}, \label{eq:colfreq}
\end{equation}
where $T$ is the temperature, $m_{\rm p}$ is the proton mass, $k_{\rm B}$ is the Boltzmann constant,  and $\sigma_{\rm in}$ is the collisional cross section ($\sigma_{\rm in} \approx 5 \times 10^{-19}$~m$^2$ for hydrogen).  In Equation~(\ref{eq:colfreq}) a strong thermal coupling between the species is assumed so that the temperature of both ions and neutrals is the same. Finally, we define the dimensional KHI timescale, $\tau_{\rm KHI}$, as
\begin{equation}
 \tau_{\rm KHI} = \frac{1}{\gamma_{\rm KHI}},
\end{equation}
where now $\gamma_{\rm KHI}$ is the dimensional growth rate. In our analysis, we have restricted ourselves to the linear stage of the KHI. This stage represents the initial phase in which the unstable mode grows after its excitation. Plasma perturbations become important during the subsequent nonlinear evolution of the KHI. However, the growth rate of the linear phase provides us with a reasonable estimation of the typical timescale on which the KHI would be observable.

\subsection{Turbulent plumes}

 \citet{berger08,berger10,berger11} and \citet{ryutova} reported on the existence of turbulent flows and  instabilities in quiescent prominences. The observations have been interpreted in terms of Rayleigh-Taylor and Kelvin-Helmholtz instabilities \citep{berger10,ryutova}. Here we are interested in the observations of turbulent flows called {\em dark plumes} \citep[see a detailed explanation of the phenomenon in][]{berger10}.

As reported by \citet{berger10}, dark plumes are turbulent upflows in prominences which usually develop Kelvin-Helmholtz vortex rolls. In Ca II H-line, plumes are seen dark in contrast to the prominence material, which suggests that the plasma in the plumes is hotter and probably less dense than the prominence material. The width of the plumes ranges between 0.5~Mm to 6~Mm and their maximum heights are between 11~Mm and 17~Mm. The mean ascent speed is about 15~km~s$^{-1}$ although velocities up to  30~km~s$^{-1}$ are measured. The typical plume lifetime is between 400~s and 890~s.

In the following analysis, medium 1 represents the prominence plasma and medium 2 the plume. We express all quantities in dimensional units. We use typical parameters of quiescent prominences: $B_{\rm 1} = B_{\rm 2} = 5$~G, $\rho_1=5\times 10^{-11}$~kg~m$^{-3}$, and $T_{\rm 1} = 10^4$~K. The exact ionization degree of the prominence plasma is unknown  \citep[see recent estimations in, e.g.,][]{labrosse,labrossereview}. We take $\zeta_{\rm n 1} = 0.5$, meaning that half of the prominence material is neutral. For the plume density, we take a contrast of 10 with respect to the prominence plasma, i.e., $\rho_2 = \rho_1 / 10$. The plume temperature, $T_2$, is computed from the pressure balance condition, which gives $T_2 = 10^5$~K. At the plume temperature we assume the plasma is largely ionized and take $\zeta_{\rm n 2} = 0.1$.  For simplicity, we assume the prominence plasma is at rest, i.e., $V_1 = 0$, and use the plume velocity, $V_2 = V_{\rm plume}$, as a parameter. We relate $L_y$ to the plume width. According to the observations \citep[see][]{berger10}, the width of the plumes is between 0.5~Mm to 6~Mm. For the present application we take $L_y = 1$~Mm. $L_z$ is more difficult to estimate from the observations, but according to the size of the vortex rolls developed in plumes, $L_z$ is probably of the same order or smaller than $L_y$. We use $L_z$ as a parameter.

\begin{figure}[!t]
\centering
\includegraphics[width=0.75\columnwidth]{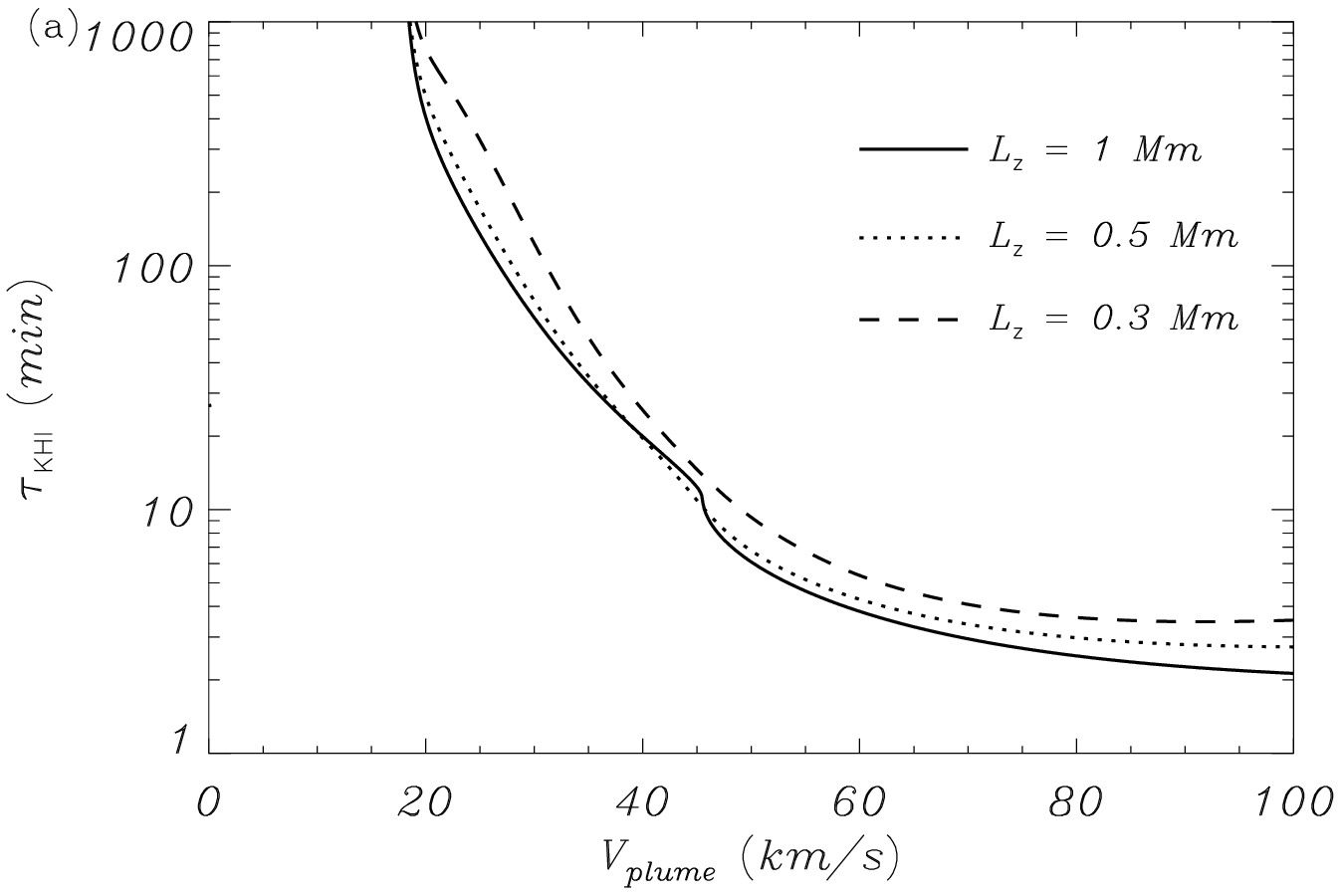}
\includegraphics[width=0.75\columnwidth]{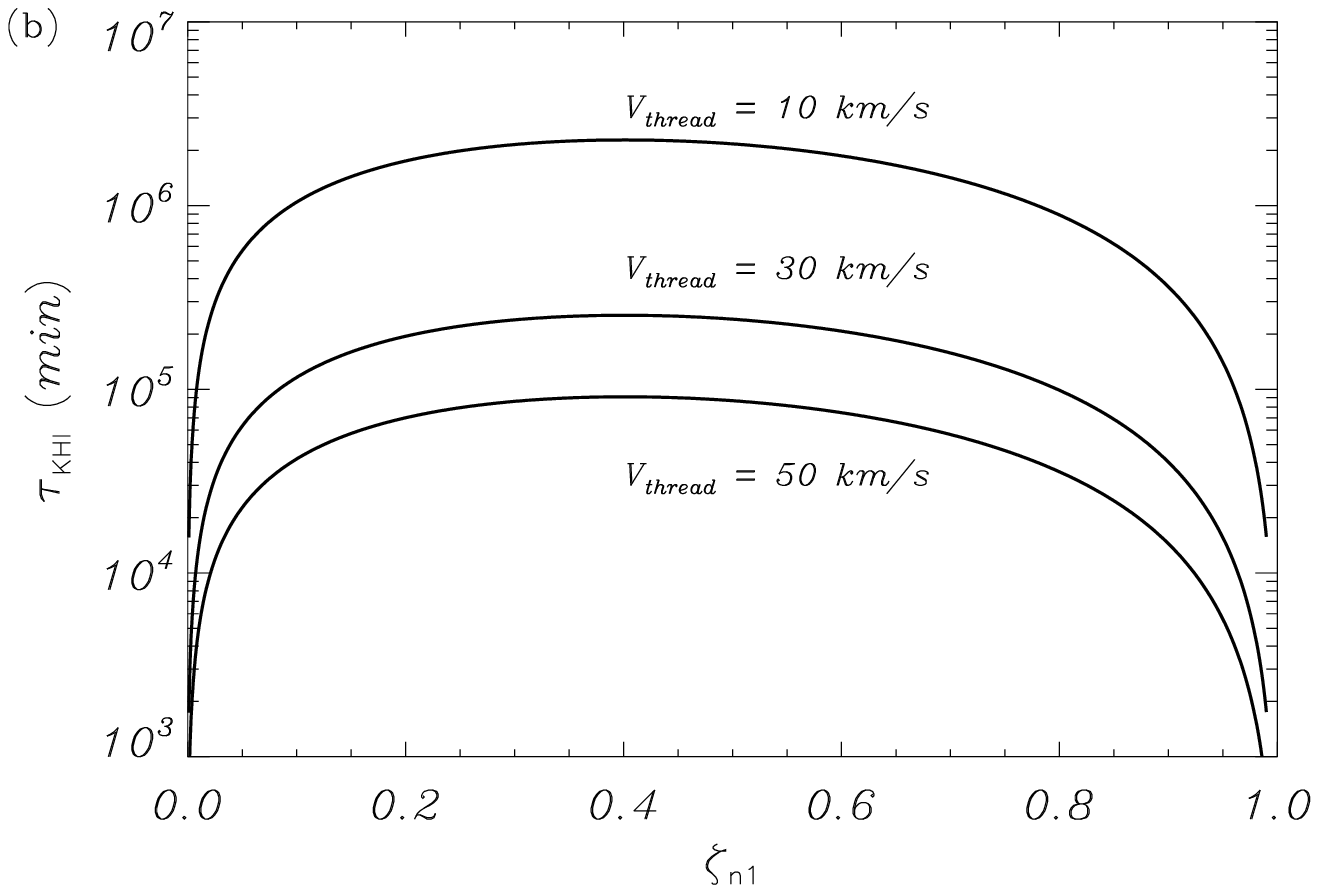}
\caption{(a) Application to turbulent plumes. KHI timescale, $\tau_{\rm KHI}$, versus plume velocity, $V_{\rm plume}$. The different lines correspond to different values of $L_z$ indicated within the Figure. (b) Application to thread in barbs. $\tau_{\rm KHI}$ versus the prominence thread ionization degree, $\zeta_{\rm n 1}$, for different values of $V_{\rm thread}$. in both cases, the remaining parameters are given in the text.  \label{fig:plume}}
\end{figure}

We display in Figure~\ref{fig:plume}(a) the dimensional KHI timescale, $\tau_{\rm KHI}$, as a function of the plume velocity, $V_{\rm plume}$, for different values of $L_z$. We obtain that $\tau_{\rm KHI}$ increases as $L_z$ decreases. This result is consistent with the idea that the smaller $L_z$, and so the larger $k_z$, the stronger the stabilizing effect of the magnetic field. Also, we see that for observed plume velocities, i.e., $V_{\rm plume} \lesssim 30$~km~s$^{-1}$, $\tau_{\rm KHI}$ is  longer than the observed plume lifetimes. For $V_{\rm plume} > 40$~km~s$^{-1}$, approximately, $\tau_{\rm KHI}$ becomes of the same order as the observed plume lifetimes. However, flow velocities of  $40$~km~s$^{-1}$ are about two times larger than observed plume velocities. 

The fact that we have to take flow velocties about two times larger than the observed ones to obtain realistic growth rates is quite a reasonable result considering the simplicity of our model. We must be aware that the present configuration, i.e., a magnetic interface, might miss relevant effects for plume dynamics. For example, body modes are not described in our model and may be important for the KHI. To take body modes into account we should consider slab or cylindrical models. This is a task to do in forthcoming works after this first investigation. In addition,  several effects not considered here may reduce the KHI timescale for smaller velocities. For example, to take the flow of neutrals not strictly parallel to the magnetic field direction may decrease the KHI timescale to values compatible with the observed instability and plume lifetimes \citep[see][]{prialnik,comets}. A different magnetic configuration may also influence the KHI growth rates.  However, the important result we want to stress here is that the KHI is present for flow velocities which are of the same order as those observed. This is due to the neutral-ion coupling, since in our configuration a fully ionized plasma would be stable for such velocities.

\subsection{Threads in prominence barbs}

High resolution observations show that prominences are form by a myriad of thin and long field-aligned ribbons, usually called {\em threads}. Threads are the basic sub-structures of prominences. They outline the magnetic field from the barbs to the spines of prominences \citep{engvold}. The observed widths of the threads are between 100~km and 600~km, while their lengths usually range between 3,000~km and 28,000~km, respectively.  Mass flows along threads have been frequently reported \citep[see the recent review by][]{labrossereview}.  The typical flow velocities are less than 30~km~s$^{-1}$ in quiescent prominences, although larger values up to 40--50~km~s$^{-1}$ have been observed in active region prominences. A recent review on the properties of prominence threads can be found in \citet{linrev}.

The lifetime of threads in H$\alpha$ observations is about 10--20 minutes, before they eventually disappear \citep[e.g.,][]{lin04,lin05,lin09}. The cause of this disappearance is unknown, although several explanations invoking instabilities have been proposed \citep[e.g.,][]{temuryKHI,solerthermal}. In this application, medium 1 represents the thread and medium 2 the inter-thread plasma. We use $B_{\rm 1} = B_{\rm 2} = 5$~G, $\rho_1=5\times 10^{-11}$~kg~m$^{-3}$, $T_{\rm 1} = 10^4$~K, $\rho_2 = \rho_1 / 100$, and $\zeta_{\rm n 2}  =0.5$. Here we use the prominence ionization degree, $\zeta_{\rm n 1}$, as a free parameter. The inter-thread temperature, $T_2$, is computed from the pressure balance condition. According to the observed widths and lengths of the threads \citep[see, e.g.,][]{linrev}, we take $L_y = 200$~km and $L_z = 10,000$~km. Hence, $k_y / k_z = 50$, meaning that we are in the incompressible regime. For simplicity, we assume the inter-thread plasma is at rest, i.e., $V_2 = 0$, and use the flow velocity in the thread, $V_1 = V_{\rm thread}$, as a parameter.

The behavior of the KHI timescale, $\tau_{\rm KHI}$, with $V_{\rm thread}$ is similar to that shown in Figure~\ref{fig:plume}(a) for the case of plumes. However, the values of $\tau_{\rm KHI}$ are significantly larger in the case of thread conditions. We display in Figure~\ref{fig:plume}(b) $\tau_{\rm KHI}$ as a function of the thread ionization degree, $\zeta_{\rm n 1}$, for different values of $V_{\rm thread}$. The ionization degree has little impact on $\tau_{\rm KHI}$ unless we approach the fully ionized ($\zeta_{\rm n 1} \to 0$) or the neutral  ($\zeta_{\rm n 1} \to 1$) limits, in which the timescale decreases. The very large values of $\tau_{\rm KHI}$ obtained indicate that, although the KHI is present, it can hardly be responsible for the instability of threads. This result encourages us to pursue other explanations in the future.

\section{Conclusion}
\label{sec:con}

In this paper we have studied the KHI due to shear flow at an interface between two partially ionized plasmas. We have focused on the effects of ion-neutral collisions and compressibility. Our results are summarized as follows. 

In a collisionless plasma, neutrals are unstable for any velocity shear, while the ion-electron fluid is  unstable for super-Alfv\'enic shear only thanks to the stabilizing role of the magnetic field. When ion-neutral collisions are included, ions and neutrals get coupled. In the incompressible case, we obtain that ion-neutral collisions are not able to stabilize neutrals regardless the value of the collision frequency. The KHI is present for any velocity shear.  Hence, our results confirm the previous findings by \citet{watson}. In the compressible case, the domain of instability depends strongly on the plasma parameters. In particular, the values of the collision frequency and the density contrast are relevant. For high collision frequencies and low density contrasts neutrals become stable and the KHI can be suppressed for sub-Alfv\'enic velocity shear. For high density contrast neutrals are more unstable and the KHI can be present for sub-Alfv\'enic velocity shear.

We have applied our results to solar prominences. In the case of turbulent plumes, we find that the KHI is present for flow velocities of the order of those observed. The theoretical KHI timescales are larger than the values consistent with the observations, which is probably a consequence of using a  simplified model. On the contrary, in the case of flow along threads we find that the KHI cannot be responsible for the thread disappearance, so another mechanism must be involved.

Here we have restricted ourselves to the linear phase of the KHI.  Linear growth rates provide us with characteristic timescales for the instability to operate. However, nonlinear studies are needed to assess the real impact of the instability on the evolution of the plasma parameters. In a recent work \citet{jones} investigated the nonlinear evolution of the KHI in a partially ionized plasma in the ambipolar-dominated regime. Despite the differences in set-up and formalism, it is instructive to compare the results of \citet{jones} with our present findings. They used a two-dimensional configuration in the $xy$-plane, with the flow and the magnetic field parallel to their $y$-direction, and the interface parallel to the $x=0$ plane. No dependence of the perturbations in their $z$-direction was considered. Therefore, the equilibrium of \citet{jones} is equivalent to our model with our $k_y = 0$ and our $k_z$ set equal to their $k_y$.  \citet{jones} concluded that the linear growth rates in the ambipolar-dominated regime are the same as in the ideal case, i.e., in the absence of ambipolar diffusion. In our formalism the ambipolar-dominated regime in which ions and neutrals are strongly coupled is equivalent to our case $\nu_{\rm in} / \omega_{\rm k} \gg 1$. We find that the linear growth rate and the threshold shear in the limit $\nu_{\rm in} / \omega_{\rm k} \to \infty$ (see Section~\ref{sec:comp})  are the same as in the fully ionized case  but replacing the density of ions by the sum of densities of ions and neutrals. Hence, our present results and the conclusions of \citet{jones} seem to be in agreement in this particular limit. Our results predict that the growth rate depends on $\nu_{\rm in} / \omega_{\rm k}$ when we depart from the limit $\nu_{\rm in} / \omega_{\rm k} \to \infty$.

The present investigation may be extended in the future by taking into account different flow velocities and directions for ions and neutrals, by incorporating the effect of gravity, and by including nonideal effects such as nonadiabatic mechanisms or magnetic diffusion. Also, the study of the nonlinear evolution of the KHI for arbitrary collision frequency is of  interest.

 \acknowledgements{
  We thank the unknown referee for helpful comments and suggestions. RS thanks Marc Carbonell for advise on the numerical solutions of the dispersion relations. RS and JLB acknowledge discussion within ISSI Team on ``Solar Prominence Formation and Equilibrium: New data, new models''. All the authors acknowledge the support from the Spanish MICINN through project AYA2011-22846. RS acknowledges support from a Marie Curie Intra-European Fellowship within the European Commission 7th Framework Program  (PIEF-GA-2010-274716). AJD acknowledges support from the MICNN through project AYA2010-1802. MG acknowledges support from K.U. Leuven via GOA/2009-009. RS and JLB also thank the financial support  from CAIB through the ``Grups Competitius'' scheme.}

\appendix

\section{Dispersion relation in the compressible collisional case}
\label{app}

In the compressible collisional case, the dispersion relation is $\mathcal{D} \left( \omega \right)  = 0 $, with $\mathcal{D} \left( \omega \right)$ given by the solution of the following determinant,
\begin{equation}
\mathcal{D} \left( \omega \right)  = \left|  
\begin{array}{cccc}
 a_{11} & a_{12}  & a_{13} & a_{14}  \\
 a_{21} & a_{22}  & a_{23} & a_{24}  \\
 a_{31} & a_{32}  & a_{33} & a_{34}  \\
 a_{41} & a_{42}  & a_{43} & a_{44}  
\end{array}
\right|,  \label{eq:genreldisper}
\end{equation}
with
\begin{eqnarray}
 a_{11} &=& C_{1,1} + C_{2,1} C_{3,1} \left(k^2_{+1} - k^2_{\rm \perp n 1} \right), \\
a_{12} &=& C_{1,1} + C_{2,1} C_{3,1} \left(k^2_{-1} - k^2_{\rm \perp n 1} \right), \\
a_{13} &=& -C_{1,2} - C_{2,2} C_{3,2} \left(k^2_{+2} - k^2_{\rm \perp n 2} \right), \\
a_{14} &=& -C_{1,2} - C_{2,2} C_{3,2}  \left(k^2_{-2} - k^2_{\rm \perp n 2} \right), \\
a_{21} &=& C_{4,1}, \\
a_{22} &=& C_{4,1}, \\
a_{23} &=& -C_{4,2}, \\
a_{24} &=& -C_{4,2}, \\
a_{31} &=& C_{5,1} k_{+1} + C_{3,1} C_{6,1} k_{+1} \left(k^2_{+1} - k^2_{\rm \perp n 1} \right), \\
a_{32} &=& C_{5,1} k_{-1} + C_{3,1} C_{6,1} k_{-1} \left(k^2_{-1} - k^2_{\rm \perp n 1} \right) , \\
a_{33} &=& C_{5,2} k_{+2} + C_{3,2} C_{6,2} k_{+2} \left(k^2_{+2} - k^2_{\rm \perp n 2} \right), \\
a_{34} &=& C_{5,2} k_{-2} + C_{3,2} C_{6,2} k_{-2} \left(k^2_{-2} - k^2_{\rm \perp n 2} \right), \\
a_{41} &=& \left( C_{5,1}  C_{7,1} + C_{8,1}  \right) k_{+1} + C_{3,1} C_{5,1} C_{6,1} k_{+1} \left(k^2_{+1} - k^2_{\rm \perp n 1} \right), \\
a_{42} &=& \left( C_{5,1}  C_{7,1} + C_{8,1}  \right) k_{-1} + C_{3,1} C_{5,1} C_{6,1} k_{-1} \left(k^2_{-1} - k^2_{\rm \perp n 1} \right), \\
a_{43} &=& \left( C_{5,2}  C_{7,2} + C_{8,2}  \right) k_{+2} + C_{3,2} C_{5,2} C_{6,2} k_{+2} \left(k^2_{+2} - k^2_{\rm \perp n 2} \right), \\
a_{44} &=& \left( C_{5,2}  C_{7,2} + C_{8,2}  \right) k_{-2}  + C_{3,2} C_{5,2} C_{6,2} k_{-2} \left(k^2_{-2} - k^2_{\rm \perp n 2} \right), 
\end{eqnarray}
where parameters $C_1$ -- $C_8$ are defined as
\begin{eqnarray}
 C_1 &=&  \rho_{\rm i } \frac{k_z^2 \csi^2 \vai^2}{\Omega^2} \frac{ i \nu_{\rm in}}{\Omega + i \left( 1+\chi \right) \nu_{\rm in}}, \\
C_2 &=& -\rho_{\rm i }\left( \csi^2 + \vai^2 - \frac{k_z^2 \csi^2 \vai^2}{\Omega^2} \frac{\Omega + i \nu_{\rm in}}{\Omega + i \left( 1+\chi \right) \nu_{\rm in}}   \right), \\
C_3 &=&  - i \frac{\csn^2}{\Omega \nu_{\rm in}}, \\ 
C_4 &=&  -\rho_{\rm n} {\csn^2}, \\
C_5 &=& - \frac{1}{\Omega^2} \frac{\Omega^2 \chi \csn^2  - k_z^2 \csi^2 \tilde{c}^2_{\rm Ai}}{\Omega^2 - k_z^2 \tilde{c}^2_{\rm Ai}}  \frac{i \nu_{\rm in}}{\Omega + i \left( 1+\chi \right) \nu_{\rm in}}, \\
C_6 &=& -\frac{1}{\Omega^2} \frac{\Omega^2 \left(\csi^2 + \vai^2 \right)  - k_z^2 \csi^2 \tilde{c}^2_{\rm Ai}}{\Omega^2 - k_z^2 \tilde{c}^2_{\rm Ai}}  \frac{\Omega + i \nu_{\rm in}}{\Omega + i \left( 1+\chi \right) \nu_{\rm in}}, \\ 
C_7 &=&  i  \frac{ \nu_{\rm in}}{\Omega + i  \nu_{\rm in}}, \\
C_8 &=& - \frac{\csn^2}{\Omega \left( \Omega + i \nu_{\rm in} \right)}. 
\end{eqnarray}

\end{document}